\documentclass[aps,prl,reprint,floatfix]{revtex4-2}
\pdfoutput=1

\usepackage{silence}
\usepackage[english]{babel}

\usepackage[T1]{fontenc}
\usepackage{stix2}
\usepackage{upgreek}
\newcommand{\upDelta}{\Updelta}

\usepackage[version=4]{mhchem}

\usepackage{siunitx}

\usepackage[stretch=15,shrink=15,step=1]{microtype}

\usepackage[RGB]{xcolor}

\usepackage{nicefrac}

\usepackage{checkend}

\usepackage{graphicx}

\usepackage{float}

\usepackage{booktabs}

\usepackage{mleftright}

\usepackage{fancyhdr}

\usepackage{xfp}

\usepackage{bibunits}
\usepackage{natbib}

\usepackage[bookmarks,pdftitle={Constriction of actin rings by passive crosslinkers},pdfauthor={Alexander Cumberworth, Pieter Rein ten Wolde},pdffitwindow=false,pdfstartview=FitH,pdfdisplaydoctitle,colorlinks,plainpages=false,pdfpagelabels,hypertexnames,citecolor={blue},linkcolor={teal},urlcolor={violet},pdflang={en},hyperfootnotes=false,breaklinks]{hyperref}

\usepackage[capitalise]{cleveref}

\definecolor{textblack}{gray}{0.125}

\allowdisplaybreaks

\let\oldheadrule\headrule
\renewcommand{\headrule}{\color{textblack}\oldheadrule}

\fancypagestyle{main}{
    \fancyhf{}

    \fancyfoot[C]{\thepage}
    \pagenumbering{arabic}
}

\input{tex/macros}

\graphicspath{{figs-revision1/}}

\begin{document}
    \title{\textcolor{textblack}{Constriction of actin rings by passive crosslinkers}}
    \author{\textcolor{textblack}{Alexander Cumberworth}}
    \email{alex@cumberworth.org}
    \author{\textcolor{textblack}{Pieter Rein ten Wolde}}
    \email{tenwolde@amolf.nl}
    \affiliation{\textcolor{textblack}{AMOLF, Science Park 104, 1098 XG Amsterdam, Netherlands}}

    \begin{abstract}
        \defaultcolor{textblack}
        In many organisms, cell division is driven by the constriction of a cytokinetic ring, which consists of actin filaments and crosslinking proteins.
While it has long been believed that the constriction is driven by motor proteins, it has recently been discovered that passive crosslinkers that do not turn over fuel are able to generate enough force to constrict actin filament rings.
To study the ring constriction dynamics, we develop a model that includes the driving force of crosslinker condensation and the opposing forces of friction and filament bending.
We analyze the constriction force as a function of ring topology and crosslinker concentration, and predict forces that are sufficient to constrict an unadorned plasma membrane.
Our model also predicts that actin-filament sliding arises from an interplay between filament rotation and crosslinker hopping, producing frictional forces that are low compared to those of crosslinker-mediated microtubule sliding.

    \end{abstract}

    \maketitle
    \defaultcolor{textblack}
    \pagestyle{main}
    In animals, fungi and some closely related unicellular eukaryotes, cytokinesis is driven by the assembly and constriction of a ring composed of actin filaments and a number of actin-binding proteins~\cite{pollard2019,mangione2019,cheffings2016}.
It is widely believed that the constriction forces in the ring are generated by myosin, a motor protein that drives the sliding of actin filaments along each other in an active process using fuel turnover~\cite{mcdargh2021,nguyen2018,thiyagarajan2017,murrell2015,oelz2015,turlier2014,jung2014}.
However, ring constriction and successful cell division have been observed with impaired myosin motor activity~\cite{ma2012}, particularly in budding yeast~\cite{mendes-pinto2013,mendes-pinto2012,lord2005}.
Furthermore, it was recently found that passive crosslinking proteins, which do not turnover fuel, are also able to produce sliding forces, either via the entropy associated with the diffusion of the crosslinkers within the overlap region between the filaments, or via their condensation from the solution to the overlap region~\cite{lansky2015}. 
Even more recently, it has been shown that passive actin crosslinkers can not only induce the assembly of actin filaments into rings, but also even drive the constriction of these rings~\cite{kucera2021}.
Yet, it remains unclear what the magnitude of the constriction force is that can be generated via this mechanism, how this force depends on the concentration of the crosslinkers and the stiffness and configuration of the filaments, and whether these forces would be sufficient to drive cell constriction.
Finally, given that microtubule sliding driven by passive crosslinkers rapidly stalls because the friction becomes prohibitive~\cite{lansky2015,wierenga2020}, it is also unclear how actin rings are able to constrict on experimental timescales~\cite{kucera2021}.

Here, we develop an analytical model that accounts for the sliding force from passive filament crosslinkers and the opposing forces stemming from the bending of filaments and the friction of sliding.
The details of our model are based on the anillin-actin system of Ref. \cite{kucera2021}, although we emphasize that the framework of the model is more generic, applying to any system consisting of filaments that are crosslinked by proteins that passively bind to discrete binding sites.
We show under what conditions a ring at equilibrium is possible, and how the force depends on the ring radius for different filament lengths, topologies, and crosslinker concentrations.
The force generated by such a ring would be sufficient to constrict an unadorned plasma membrane, and could be of interest for building the division machinery in a minimal synthetic cell.
Lastly, our model predicts that the frictional force scales exponentially with the number of bound crosslinkers, in contrast with the superexponential scaling of crosslinker-mediated microtubule sliding~\cite{wierenga2020}, allowing for ring constriction on minute timescales~\cite{kucera2021}.

In order to model the contractile rings, we need an expression for the energy of filament sliding when passive crosslinkers are present.
In this case, the sliding force is a condensation force, i.e.~the sliding is driven by the binding of more crosslinkers as the overlap increases~\cite{lansky2015}.
Both \citet{lansky2015} and \citet{wierenga2021} derived analytical expressions for a condensation force where the binding and unbinding of the crosslinkers were assumed to be fast relative to filament sliding.
Here, as in Ref.~\cite{wierenga2021}, we assume that the crosslinkers can either bind once to a single filament or to a pair of binding sites, such that they bundle two filaments (see Supplemental Material~\cite{Note1} for discussion of anillin-actin bundling).
If the binding sites are separated by a distance of $\deltad$, then the overlap length is $L = \deltad (l - 1)$, where $l$ is the number of binding site pairs (\cref{fig:diagram}(a)).
The free energy is then given by~\cite{wierenga2021}
\begin{equation}
    \upDelta \Phi_\mathrm{S} = -\frac{\kb T L}{\deltad} \ln \mleft( 1 + \xi \mright),\quad \xi = \frac{{\Ks}^2 \ce{[X]}}{\Kd (\Ks + \ce{[X]})^2},
    \label{eq:free-energy}
\end{equation}
where $T$ is the temperature, $\ce{[X]}$ is the concentration of anillin, $\Ks$ is the dissociation constant of anillin binding to a single actin filament, $\Kd$ is the dissociation constant of anillin crosslinking two actin filaments from solution, and $\kb$ is Boltzmann's constant.
The linear dependence of $\upDelta \Phi_\mathrm{S}$ on $L$ means that the sliding force, given by the derivative of \cref{eq:free-energy}, does not depend on the overlap length.

Because the sliding force is independent of the overlap length, the ring-constriction force depends only on the total number of overlaps, and hence the connectivity of the filaments, or in other words, the topology of the ring.
We consider stable rings wherein all filaments are under tension and hence contribute to the constriction force (i.e.~those filaments that have overlaps on either end; see Supplemental Material~\cite{Note1} for further discussion).
To characterize the ring topology, we introduce the notion of a ``scaffold ring'' (\namecrefs{fig:diagram} \labelcref{fig:diagram}(b), \labelcref{fig:energy-force}(a), and \ref{fig:topology-diagrams-sup}(a)).
The scaffold ring is defined by the smallest set of filaments that still form a ring, and $\Nsca$ is defined to be the number of filaments in that ring.
Making the simplifying assumption that all filaments have the same length, the number of overlaps in the ring is determined by $\Nsca$ and the total number of filaments in the ring $\Nf$ (\cref{fig:diagram}(b)).
In a scaffold ring, each filament forms two overlaps, making the number of overlaps equal to the number of filaments.
However, each additional filament will add two more overlaps to the ring; the total number of overlaps is thus $\No = 2\Nf - \Nsca$.
One could envision the scaffold ring as the ring that initially forms, with subsequent binding of filaments leading to rings with $\Nf > \Nsca$.

\begin{figure}
    \includegraphics{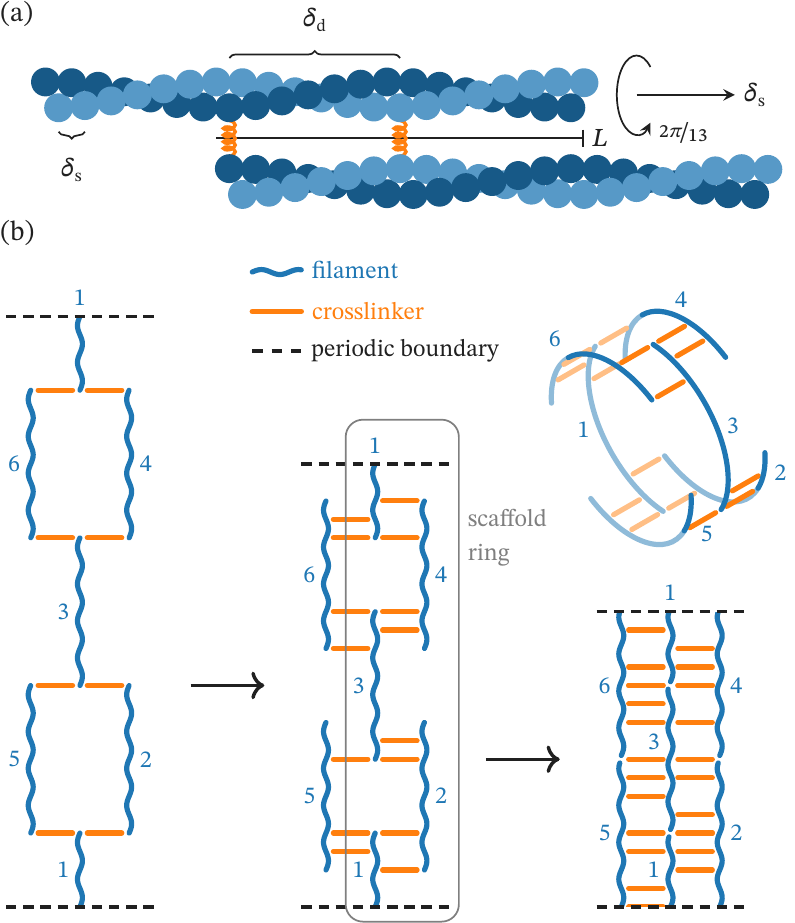}
    \caption{
        \label{fig:diagram}
        (a) Diagram of a single overlap of length $L$ formed by two actin filaments, with two of the three crosslinking sites filled.
        The half helical pitch determines the distance between crosslinking sites $\deltad$ rather than the monomer-monomer distance $\deltas$.
        Filaments slide in increments of $\deltas$, which requires a rotation of $\nicefrac{2 \pi}{13}$.
        (b) Diagrams of a ring constricting to half its radius.
        The 2D diagrams use periodic boundary conditions.
        On the left, the ring is at its maximum radius $\Rmax$; in the middle, it has constricted to three quarters of $\Rmax$; on the right, it has reached its minimum radius $\Rmin = \nicefrac{\Rmax}{2}$.
        In the top right, a 3D representation of the topology diagram in the center is given.
        For a given $\Nf$, the topology of the ring is characterized by the ``scaffold ring''.
        In this example, the number of filaments in the scaffold ring $\Nsca$ is 4, while the total number of filaments $\Nf$ is 6.
        Note that the scaffold ring is not unique; here the minimal set is defined by filaments 1, 2, 3, and 4, but it could just as well be defined by 1, 3, 5, and 6.
     }
\end{figure}

To derive the total free energy of the ring as a function of the radius, we need an expression for the total overlap length $\Ltot$ as a function of the radius $R$.
If we make the further simplifying assumption that all filaments overlap to the same degree, then $\Nsca$ also determines this relationship: the length of each overlap $L$ is the difference between the filament length $\Lf$ and the current ring circumference divided by $\Nsca$.
Then, the total overlap length is
\begin{equation}
    \Ltot = \No L = ( 2 \Nf - \Nsca ) \mleft( \Lf - \frac{2 \pi R}{\Nsca} \mright).
    \label{eq:connector}
\end{equation}
This expression reveals that the maximum radius $\Rmax$, where $\Ltot = 0$, is $\Rmax = \nicefrac{\Nsca \Lf}{2 \pi}$.
We restrict the minimum radius $\Rmin$ to be $\nicefrac{\Rmax}{2}$ because of an energy barrier due to steric clashing that begins to occur at this point as the filaments run into each other, a point we explain in more detail in the Supplementary Material~\cite{Note1}.

The final element in deriving an expression for the constriction force is an expression for the energy required to bend the filaments into a ring.
If we treat the ring as a simple linearly elastic rod, and assume that the ring geometry is perfectly circular (the full derivation and justifications for these assumptions are given in the Supplemental Material~
\footnote{See Supplemental Material for discussion of the model parameters and configuration space assumptions, derivations of the ring bending energy and friction coefficients, and supplemental figures.}%
), then the bending energy is
\begin{equation}
    U_\mathrm{B} = \frac{\Nf E I \Lf}{2 R^2}, \label{eq:bending-energy-main}
\end{equation}
where $E$ is the Young's modulus, $I$ is the second moment of area, and $R$ is the ring radius.
Then, multiplying \cref{eq:free-energy} by $\No$ and combining it with \cref{eq:bending-energy-main,eq:connector} allows us to write the total free energy, the derivative of which gives the radial constriction force,
\begin{equation}
    F = -\frac{2 \pi \kb T (2 \Nf - \Nsca)}{\Nsca \deltad} \ln \mleft( 1 + \xi \mright) + \frac{\Nf EI \Lf}{R^3}.
    \label{eq:tracks-force}
\end{equation}
In equilibrium, the sliding force is balanced by the opposing bending force such that the net force $F$ is zero; solving \cref{eq:tracks-force} for $F=0$ then yields the equilibrium radius $\Req$.

\begin{figure}
    \includegraphics{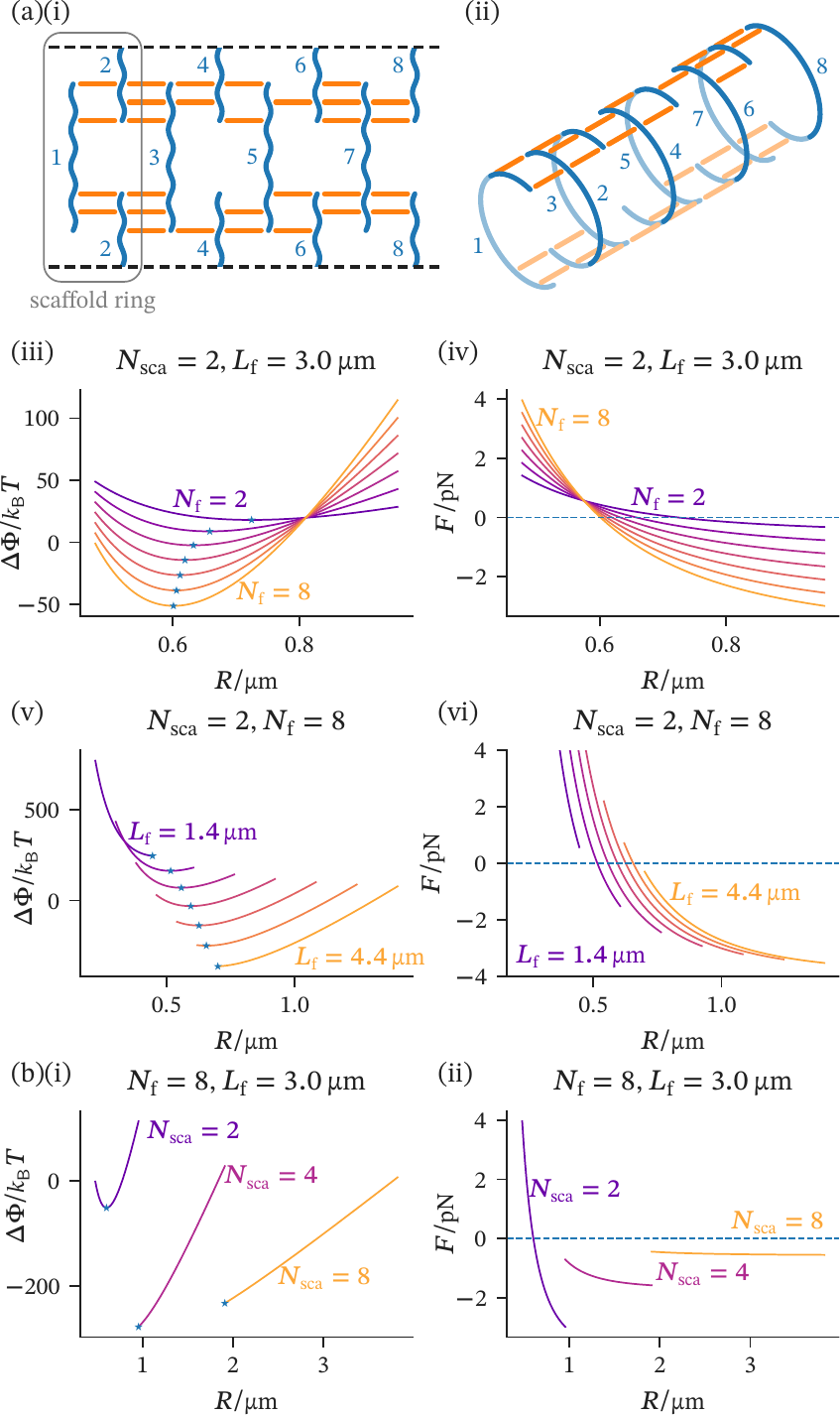}
    \caption{
        Total free energy and total force as a function of the ring radius.
        The blue stars indicate the points at which the free energy is minimal, and negative force values indicate a constricting force.
        In (a), $\Nsca = 2$, as per the topology depicted with both 2D periodic boundary conditions (i) and in 3D (ii); the free energies and total forces are then plotted for either a range of $\Nf$ values ((iii) and (iv)) or a range of $\Lf$ values ((v) and (vi)).
        The constriction force increases linearly with $\Nf$.
        In (b), $\Nf = 8$ for three values of $\Nsca$.
        Connecting the filaments in series (large $\Nsca$) increases the constriction range, while stacking them in parallel (small $\Nsca$) increases the maximum constriction force.
        Data for all plots was produced with the parameters given in Table S1~\cite{Note1}.
        \label{fig:energy-force}
     }
\end{figure}

With \cref{eq:tracks-force}, we can calculate the free energies and constriction forces as a function of the radius, as well as the equilibrium radius for a given set of parameters.
In the study that observed the constricting rings~\cite{kucera2021}, it was found that the rings have up to eight filaments, so we focus on this range of filaments here (see Supplemental Material~\cite{Note1} for discussion of model parameters).
At large radii, the sliding force dominates over the bending force, and the net force is negative (inward) (\cref{fig:energy-force}).
The free energy and the magnitude of the force then decrease as the radius decreases.
For radii smaller than $\Req$, the free energy increases and the net force becomes positive (outward), since the bending force becomes larger than the sliding force.
As $\Nf$ increases for a given $\Nsca$, $\Req$ decreases, but quickly enters the regime of heavily diminishing returns (\cref{fig:energy-force}(a)(iii)).
From \cref{fig:energy-force}(a)(iv) and \cref{eq:tracks-force}, however, the initial constriction force can be seen to increase linearly with $\Nf$ for a given radius.
Finally, for a given $\Nf$, there is a trade-off between the range of radii over which the ring can generate force and the maximum possible force it can achieve.
While connecting the filaments in series (large $\Nsca$) favours the former, stacking them in parallel (small $\Nsca$) favours the latter (\cref{fig:energy-force}(b)).

As can be seen from \cref{eq:tracks-force}, there is an optimal anillin concentration that maximizes the force, specifically when $\ce{[X]} = \Ks$, after which point higher concentrations lead to lower sliding forces~\cite{wierenga2021}.
This decrease in the sliding force occurs because, at higher crosslinker concentrations, the entropy of mixing in solution favours the binding of crosslinkers to a single actin site instead of two, thus impeding crosslinking and hence the condensation force.
In fact, the anillin concentration used in the experiments~\cite{kucera2021} happens to be close to $\Ks$ (\namecref{fig:diagram} \ref{fig:force-concentration}).

The ring-constriction experiments of \citet{kucera2021} were performed under three conditions: stabilized, non-stabilized, or destabilized against depolymerization.
Interestingly, the constriction of the rings differed markedly under these conditions.
While differences in the stabilized and non-stabilized conditions could be explained by the difference in the bending rigidity (the stabilizer rigidifies actin~\cite{isambert1995}), another explanation is needed for the observation that the destabilized conditions led to further constriction than the other two conditions.
Our model provides an explanation for this observation: as the filaments depolymerize, the bending force required for a given radius is lower, and so the equilibrium radius will be smaller.
If one considers a point on one of the longer filament length curves in \cref{fig:energy-force}(a)(v), disassembly can be thought of as moving down to another curve of a smaller filament length.

So far, we have ignored the frictional forces involved in ring constriction.
To study the timescale of ring constriction, we simulate the overdamped ring-constriction dynamics within our model.
To do so, we require an expression for the friction coefficient of the ring.

Wierenga and Ten Wolde~\cite{wierenga2020} derived the friction coefficient of the sliding of two crosslinked microtubules.
The sliding mechanism involves a collective rearrangement of all the crosslinkers in the overlap, wherein the crosslinkers hop between neighbouring sites on the microtubules.
This collective rearrangement gives rise to an energy barrier that increases linearly with the number of bound crosslinkers $\Nd$, and thus a friction coefficient that scales exponentially with $\Nd$.
Additionally, because the crosslinkers can block each other from hopping, the friction scales superexponentially with $\Nd$ at higher crosslinker densities.

In actin filaments, the distance between crosslinker binding sites $\deltad$ is much larger than the distance over which crosslinkers can hop, as the distance between sites is set by the helical pitch rather than the monomer-monomer distance $\deltas$~\cite{matsuda2020} (\cref{fig:diagram}(a)).
This suggests that the above sliding mechanism may be unfeasible.
However, actin filaments have been observed to twirl, or rotate, about their axes when they are being slid by myosin motor proteins bound to a surface~\cite{jegou2020,ropars2016,vilfan2009,beausang2008,ali2002,sase1997}.
We postulate that passively crosslinked actin filaments must also twirl as they slide, which would allow crosslinkers to hop distances of $\deltas$ (rather than the larger $\deltad$), thus also allowing the same collective rearrangement as was found for microtubules (\namecrefs{fig:diagram} \labelcref{fig:diagram}(a) and \ref{fig:sliding}).

Because the crosslinkers are spaced at least $\deltad$ apart, they do not block each other when hopping by $\deltas$ to neighbouring sites (\cref{fig:diagram}(a)).
The friction coefficient therefore only scales exponentially with $\Nd$, not superexponentially as with microtubule sliding~\cite{wierenga2020}; as we will see, this difference has dramatic consequences.
In the regime where crosslinker binding is fast relative to filament sliding and the average number of crosslinkers is a function of the overlap $L$, the friction coefficient of a single overlap is
\begin{equation}
    \zeta = \zeta_0 \mleft[ \frac{1 + \xi}{1 + \xi \expsup{-B}} \mright]^{\mleft( \frac{L}{\deltad} + 1 \mright)} \label{eq:zeta}.
\end{equation}
Here, $\xi$ is defined in \cref{eq:free-energy} and
\begin{equation}
    \zeta_0 = \frac{\kb T}{\deltas^2 r_0} \sqrt{1 + \frac{3 k \deltas^2}{4 \kb T}},\quad B = \frac{\deltas^2 k}{8 \kb T} - \ln 2,
\end{equation}
where $k$ is the spring constant of the crosslinker, and $r_0$ is the jump-rate prefactor for crosslinker hopping (see Supplementary Information~\cite{Note1} for full derivation).
The friction coefficient of the ring $\zetaR$ can be related to that of a single overlap $\zeta$ via $\zetaR = \nicefrac{4 \pi^2 \No}{\Nsca^2} \zeta$ by considering the numbers of overlaps acting in series and in parallel (see Supplementary Information~\cite{Note1} for full derivation).

Simulations with this expression for the friction coefficient show that the ring constriction occurs within experimental timescales~\cite{kucera2021} for the experimentally relevant range of model parameter values (\cref{fig:dynamics}).
Of the parameters, the spring constant has been shown to be the most critical~\cite{lansky2015,wierenga2020}.
While the value of $k$ for anillin has not been measured, crosslinking proteins tend to be on the lower end of the range of values measured in proteins~\cite{pratap2022,nakagawa2020,wierenga2020,grimaldo2019,ahmadzadeh2014,tseng2009,dong2009,aprodu2008,kawakami2006,jeney2004,zaccai2000,duke1999}, with values falling within \SIrange[range-units=single]{0.3}{1.2}{\pico\newton\per\nano\meter}~\cite{duke1999,wierenga2020,jeney2004,ahmadzadeh2014}.
To explore the dynamics in this range of $k$, we first consider a ring with $\Nsca = 2$ and $\Nf = 5$, and a filament length of \SI{3.0}{\micro\meter}, which leads to an equilibrium radius that is around halfway between the maximum and minimum possible radii (\namecref{fig:diagram} \ref{fig:Req}(b)).
With $k = \SI{1.0}{\pico\newton\per\nano\meter}$, the ring constricts almost instantaneously (\cref{fig:dynamics}).
Increasing the value to $k = \SI{1.5}{\pico\newton\per\nano\meter}$ leads to much slower constriction, but the ring still reaches its equilibrium value within the timescale used in the experiments (\SI{1000}{\second})~\cite{kucera2021}.
By $k = \SI{2.0}{\pico\newton\per\nano\meter}$, the ring constriction stalls before reaching the equilibrium radius.
Varying $\Nf$ (\namecrefs{fig:dynamics} \ref{fig:dynamics-sup}(b) and \ref{fig:dynamics-k-2}) and using a different value of $\Lf$ (\namecref{fig:dynamics} \ref{fig:dynamics-Lf-4}) did not have a large impact on the dynamics.
Overall, while microtubule sliding rapidly stalls because the friction becomes prohibitive~\cite{lansky2015}, the much weaker scaling of the friction with $\Nd$ for anillin-actin rings means that the latter can constrict to their equilibrium radii on experimental timescales of minutes~\cite{kucera2021}.

Even when the mean forces on either end of the filaments are balanced, thermal fluctuations in the positions of the filaments imply that the ring will eventually break.
However, in Ref.~\cite{kucera2021}, the rings were observed to be stable on the timescale of the experiments.
During ring constriction, a mechanism exists that would tend to equalize overlaps: if some overlaps fluctuate towards smaller values, then the ring-constriction force will tend to increase their overlap more quickly because they also then have a lower friction.
Once the rings have reached equilibrium, we can estimate the timescale on which the ring is stable from $\tau = \nicefrac{x^2}{\DLf}$, where $\DLf$ is the diffusion coefficient of a single filament and $x$ is the overlap length at equilibrium.
The overlaps of a single filament act in parallel, so the total friction force experienced by a single filament is the sum of the friction of its $n$ individual overlaps, which implies that the friction coefficient of a single filament is $\zetaf = n \zeta$.
We can relate this friction coefficient to the diffusion coefficient of a filament with $\DLf = \nicefrac{\kb T}{\zetaf}$.
If we consider a filament with two overlaps, and use the overlap length of the rings modeled in \cref{fig:dynamics} in equilibrium, $x = \SI{0.75}{\micro\meter}$, as well as the value of the friction coefficient at equilibrium, $\zeta = \SI{5e-4}{\second\per\kilo\gram}$, then $\tau = \SI{14000}{\second}$.
This is more than an order of magnitude longer than the experimental timescale.
Further, this calculation assumes that the overlaps are equal in size, yet fluctuations away from this state will lengthen some overlaps at the expense of others, which tends to increase the friction the filament experiences.
In other words, the estimate for $\tau$ above gives a lower bound, strengthening the prediction that the rings are stable on much longer timescales than the experiments were performed on.

\begin{figure}
    \includegraphics{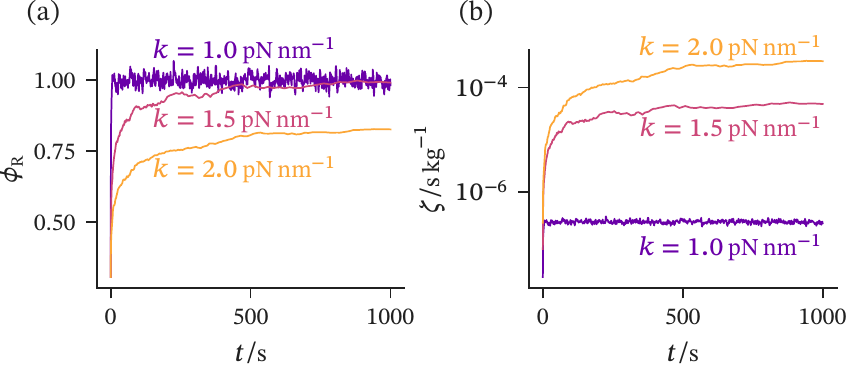}
    \caption{
        Ring-constriction trajectories with $\Lf = \SI{3.0}{\micro\meter}$, $\Nsca = 2$, $\Nf = 5$, and three different values of $k$.
        (a) Progress of radial constriction from its maximum ($\phi_\mathrm{R} = 0$) to its equilibrium value ($\phi_\mathrm{R} = 1$), where $\phi_\mathrm{R} = \nicefrac{\mleft( \Rmax - R \mright)}{\mleft( \Rmax - \Req \mright)}$.
        (b) Friction coefficient of an overlap in the ring.
        \label{fig:dynamics}
        For numerical details and additional quantities, see \namecref{fig:dynamics} \ref{fig:dynamics-sup}(a); for trajectories with values of $k$ down to \SI{0.1}{\pico\newton\per\nano\meter}, see \namecref{fig:dynamics} \ref{fig:dynamics-small-k}.
        For the biologically relevant range of $k$, constriction occurs on experimental timescales~\cite{kucera2021}.
     }
\end{figure}

The tension $T$ of cytokinetic rings, which is related to the constriction force $F$ via $T = \nicefrac{F}{2 \pi}$, has been measured to be hundreds of \si{\pico\newton} in fission yeast~\cite{mcdargh2021,stachowiak2014}, while here, using experimentally measured values of $\Nf$, $\Lf$, and $R$ in vivo~\cite{mallal2022,swulius2018,courtemanche2016}, the tension does not exceed \SI{10}{\pico\newton}.
However, even the measured ring tensions are nearly three orders of magnitude too small to balance measured turgor pressures and would, in fact, only produce strains in the cell wall of 0.01\%~\cite{pollard2019}.
Instead of directly constricting the cell, the cytokinetic ring in fission yeast coordinates and guides the inward growth of the cell wall via mechanosensitive enzymes~\cite{zhou2015b,thiyagarajan2015}.
For this function, even a much smaller ring tension may be sufficient, if less effective~\cite{thiyagarajan2015}.
These findings in fission yeast, a model organism for eukaryotic division~\cite{chang2017,fantes2016}, are likely relevant more broadly in fungi; in budding yeast, a similar interplay between cell-wall growth and cytokinetic-ring constriction has been observed~\cite{bhavsar-jog2017}.
While the maximum tension generated by passive crosslinking is smaller than that generated by myosin motors in the cytokinetic ring, the mechanism may provide an explanation for the observations that budding yeast can divide without myosin motor activity~\cite{mendes-pinto2013,mendes-pinto2012,lord2005}, possibly in combination with other non-myosin-motor force generating mechanisms~\cite{chen2020}.

Regardless of the role of passive-crosslinker-driven constriction in natural cells, the anillin-actin system is an interesting candidate for a cell-division module in a minimal synthetic cell.
Actin, under polymerizing conditions, can grow into rings centered at the equator of droplets with diameters less than the persistence length of actin~\cite{miyazaki2015}.
Anillin, besides being a crosslinker of actin, also has domains which bind to lipid membranes, and it appears to play a key role in anchoring the cytokinetic ring to the membrane in animal cells~\cite{sun2015}.
Further, experiments~\cite{schoneberg2018} and modeling~\cite{beltran-heredia2017} indicate that constriction of unadorned lipid-membrane vesicles requires contractile forces on the order of a \si{\pico\newton}, well within the range that can be generated via this mechanism.
Therefore, by coupling anillin expression to the cell cycle, the system studied here may constitute a minimal module for triggering cell division upon the completion of DNA replication~\cite{olivi2021}.

    \begin{acknowledgments}
        The authors would like to thank Vahe Galstyan and Ramon Creyghton for their helpful comments on the manuscript.
This work was supported by a Gravitation grant from the Netherlands Organization for Scientific Research (NWO) Gravitation, Building a Synthetic Cell (BaSyC) (024.003.019).

    \end{acknowledgments}
    \bibliography{tex/main.bib}
    \onecolumngrid
    \clearpage
    \pagebreak

    \twocolumngrid
    \title{\textcolor{textblack}{Supplemental Material: Constriction of actin rings by passive crosslinkers}}
    \maketitle
    \supplemental
    \begin{bibunit}[apsrev4-2]
    \makeatletter\write\@bibunitaux{\string\citation{REVTEX42Control}}\makeatother
    \makeatletter\write\@bibunitaux{\string\citation{apsrev42Control}}\makeatother
    In this Supplemental Material, we discuss the assumptions made about the model's configuration space, derive the bending energy of a ring composed of crosslinked filaments, discuss the details of the overdamped dynamics used in the ring-constriction simulations, derive the friction coefficient for ring-constriction dynamics, discuss the selection of parameters for the model, and include additional figures to supplement those in the main text.
The data, scripts, and simulation code for this work are accessible via Ref.~\cite{replication-package_sup}.

\section*{Discussion of model configuration space assumptions}

In this section, we discuss how anillin bundles actin, the definition of force-generating filaments, the conditions required for ring stability, and the justification for the assumption regarding the minimum ring radius.

\subsection*{Anillin-actin bundling}

\begin{figure*}
    \centering
    \includegraphics{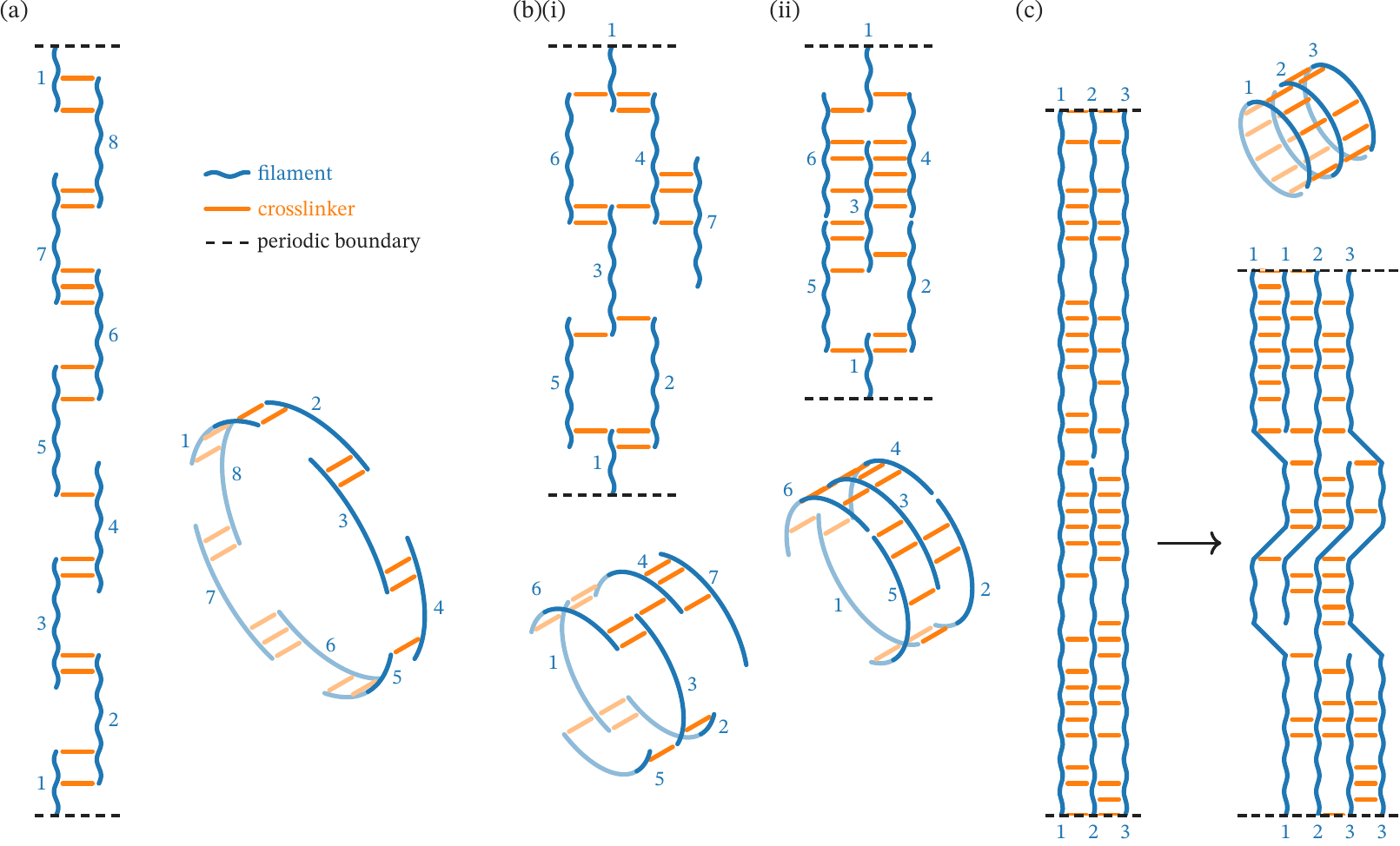}
    \caption{
        (a) Topology of a ring with $\Nf = 8$ and $\Nsca = 8$; a 2D representation with periodic boundary conditions is given on the left, while a 3D representation is given on the right.
        (b) Topology diagrams of rings with configurations not considered by the model that decrease the constriction force; diagrams with periodic boundary conditions are given at the top and 3D representations below.
        (b)(i) A ring with $\Nf = 7$ and $\Nsca = 4$ where one of the filaments (7) is bound to the ring with only a single overlap, leading to a reduction in the total constriction force that the same ring without that filament would generate; this reduction is because filament 7 produces an opposing bending force but does not contribute to the constriction force.
        (b)(ii) A ring with $\Nf = 6$ and $\Nsca = 4$ where some of the overlaps have collided, reducing the total force for constriction.
        (c) Topology diagrams of a ring with $\Nf = 3$ and $\Nsca = 2$ at and past its minimum radius.
        On the left the ring has already constricted to half its maximum radius, while on the right it has continued to constrict past this point, leading to deformation in the filaments.
        The 3D representation of the ring at its minimum radius is given on the top right; the exact placement of the crosslinkers does not correspond to those depicted on the diagram on the left, as the proportions have been changed, allowing for fewer crosslinkers to be drawn.
        \label{fig:topology-diagrams-sup}
    }
\end{figure*}

Drosphila anillin has three actin binding sites, two of which are mutually exclusive~\cite{jananji2017_sup}.
When anillin bundles actin with only the two binding sites that have the highest affinity, which are on opposing sides of anillin, the resulting bundles form single-layer sheets in which the anillin linkers are all in register~\cite{jananji2017_sup}, as in \namecrefs{fig:crosssection} \ref{fig:diagram} and \ref{fig:energy-force} of the main text, or in \cref{fig:topology-diagrams-sup,fig:crosssection}(b).
The third and weakest binding site is positioned such that it cannot form sheets if bound in combination with the other binding site it is not mutually exclusive with~\cite{jananji2017_sup}; if starting with a 2D sheet, additional filaments would have to stack on top of the sheet when binding with the third site.
In experiments, both ordered 2D sheets and thicker, less ordered bundles are observed~\cite{jananji2017_sup}.
For a small number of filaments, the configuration with the lowest free energy is the sheet formed when only the two highest affinity sites are used, as this will maximize the anillin-actin binding energy.
In our model, we assume all binding involves only these two sites.

\subsection*{Force-generating filaments}

We only consider rings where all filaments contribute to the sliding force driving constriction.
For a filament to generate force in a ring, it must have at least one overlap on each end; i.e.~the filament must be under tension.
For example, in \cref{fig:topology-diagrams-sup}(b)(i), filaments 1--6 have at least one overlap on either end, so are under tension, but filament 7 has only a single overlap, so is not under tension.
This ring would have the same sliding force driving constriction as one without filament 7, but would have a lower total constriction force because of the contribution to the bending force from the additional filament.
Since we are interested in rings that maximize the constriction force for a given number of filaments, we do not consider these configurations.

\subsection*{Ring stability}

In the analytical theory of the main text, the overlap sliding force is taken to be a condensation force, which is independent of the overlap length.
For a ring to be stable, each filament must then have an equal number of overlaps on either end, as this will balance the relative sliding forces.
Given the additional simplifying assumption that all filaments have the same length, then not all combinations of the total number of force-generating filaments $\Nf$ and the number of filaments in the scaffold ring $\Nsca$ yield rings that are stable.
More specifically, for the sliding (i.e.~condensation) forces to be balanced with each other, when $\Nsca$ is even, $\Nf - \Nsca$ must be a multiple of $\nicefrac{\Nsca}{2}$, while when $\Nsca$ is odd, $\Nf$ can only be equal to $\Nsca$.
For example, if filament 5 in \namecref{fig:crosssection} \ref{fig:diagram} of the main text were removed, the forces on filaments 1 and 3 would become unbalanced because these filaments would then have two overlaps on one end and only one overlap on the other end; this would make the ring unstable.
To be consistent, we apply these sum rules to our analysis in the main text to ensure we do not study unstable rings.

\subsection*{Equal-overlap length assumption}

\begin{figure}
    \centering
    \includegraphics{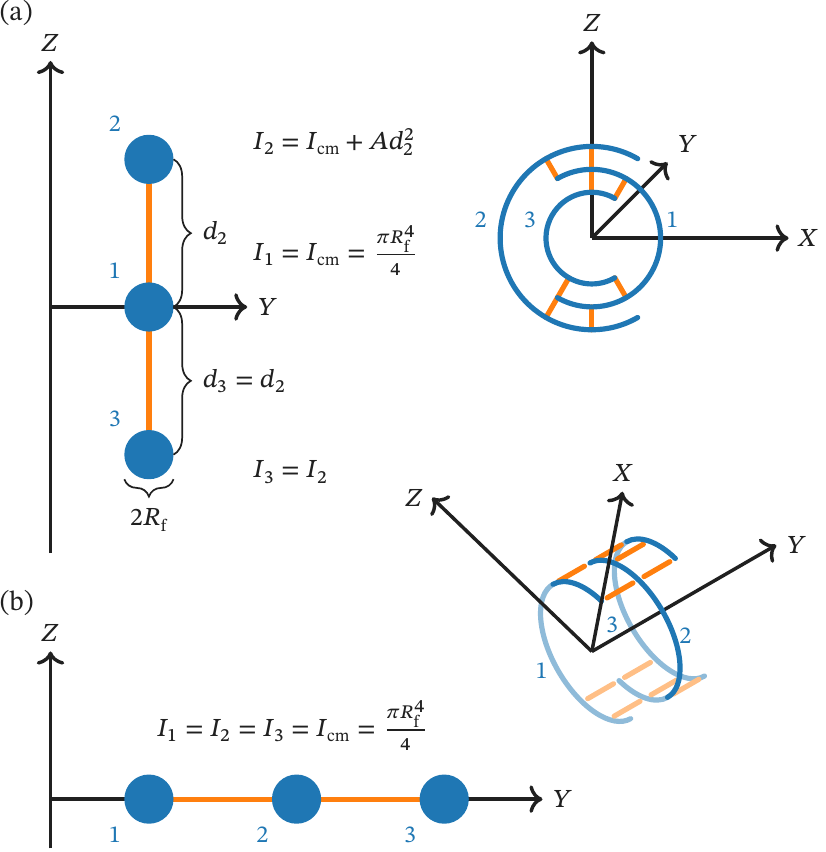}
    \caption{
        Cross sections of three crosslinked filaments to be bent around the $Y$ axis.
        (a) The three filaments are oriented such that only the middle one lies on the bending axis.
        The filaments tangentially extend in the $X$ direction, and so the cross section is drawn with the $Y$ and $Z$ axis.
        A ring configuration that would lead to this situation is given to the right.
        (b) The three filaments are oriented such that all lie on the bending axis.
        This configuration minimizes the bending energy; a ring configuration that would lead to this situation is given at the top right.
        \label{fig:crosssection}
    }
\end{figure}

In the main text, we make the simplifying assumption that all filaments overlap to the same degree.
This allows us to start from a configuration with no overlaps and follow the progress to a configuration with maximal overlaps, as in \namecref{fig:crosssection} \ref{fig:diagram}(b) of the main text.
Essentially, we consider a specific limiting case that is representative of many starting configurations.

Deviations from equal overlaps can lead to lower constriction forces, although the effect is expected to be small.
When the ring starts with overlaps that differ in size, configurations can arise where some filament ends start to collide such that their corresponding overlaps no longer drive constriction (e.g.~filaments 5 and 6 in \cref{fig:topology-diagrams-sup}(b)(ii)).
In our model, this scenario would be accompanied by a discontinuous drop in the constriction force.
If the ring starts out near its maximum radius, which is likely often the case, then the emerging filament overlaps will be fairly similar in size, making this effect small and only relevant as the ring approaches its minimum radius.
Further, the constriction dynamics will cause overlaps to tend towards being similar lengths, since larger overlaps have higher friction and will then decrease at a slower rate (see the main text for further discussion of this effect in the context of ring stability).
As such an effect does not significantly change the predictions of the model, but does increase its complexity, we do not consider it in the main text.

\subsection*{Minimum ring radius}

We consider half the maximum radius to be the minimum radius for two reasons.
First, at this point, some of the filaments would start to collide, which would lead to an energy barrier from the need to bend around each other (\cref{fig:topology-diagrams-sup}(c)).
Second, upon reaching half the maximum radius, the ability of the ring to increase its overlap is significantly reduced, as only filaments on the edge of the sheet are able to contribute to an increase in filament overlap.
To understand this, consider that the edge filaments are engaging in overlap on only one side, and so are able to add overlap on the other side (e.g.~filaments 1 and 3 in \cref{fig:topology-diagrams-sup}(c)), and hence can still sustain a sliding force.
The middle filaments, on the other hand, are engaging in overlap at all points along their length by the time they reach half the maximum radius (e.g.~filament 2 in \cref{fig:topology-diagrams-sup}(c)); any further constriction would then not change their overlap length, such that they no longer contribute to the sliding force.

\section*{Derivation of ring bending energy}

As the circumference of the rings is less than the persistence length of single actin filaments (\SI{17}{\micro\meter}~\cite{mameren2009_sup}), we cannot treat the filaments as freely jointed chains.
Moreover, multiple connected filaments will be even stiffer than single filaments.
We therefore treat the ring as a simple linearly elastic rod~\cite{arai1999_sup}.
This allows us to apply beam theory, which predicts that the elastic potential energy of a section of the ring is
\begin{equation}
    U_\mathrm{B} = \frac{1}{2} \int_0^{L_\mathrm{s}} B(s) \kappa(s)^2 \mathrm{d} s,
\end{equation}
where $B(s)$ is the bending rigidity (also known as the bending stiffness, the flexular stiffness, and the flexular rigidity), $\kappa$ is the curvature, $L_\mathrm{s}$ is the length of the section of the ring, and the integral is over the contour of the line formed by the ring with arc length parameter $s$.
If we assume the ring geometry to be perfectly circular (for anillin-actin rings this is reasonable based on the images in Ref.~\cite{kucera2021_sup}), then the curvature is simply the reciprocal of the ring radius $R$.
We can then decompose the ring into segments with different bending rigidities, based on the number of filaments and their configuration.
The bending rigidity $B = EI$ is the product of the Young's modulus $E$ with the second moment of area, $I$, which contains the geometric contribution to the bending rigidity.
The potential for segment $i$ in the ring can then be written as
\begin{equation}
    U_i^\mathrm{B} = \frac{1}{2} \int_0^{L_i^\mathrm{s}} \frac{EI_i}{R^2} \mathrm{d} s = \frac{E I_i L^\mathrm{s}_i}{2 R^2},
\end{equation}
where $L_i^\mathrm{s}$ is the length of segment $i$.

To calculate $I_i$, first consider a filament lying along the $X$ axis.
If the bending is around the $Y$ axis, then the second moment of area is defined as
\begin{equation}
    I = \int_A Y^2 \mathrm{d} A,
\end{equation}
where the integration is done over a cross section of area $A$ of the filament in an undeformed configuration.
If we assume that the filaments have cylindrical geometry, then the second moment of area for a single filament is
\begin{equation}
    I = \frac{\pi R_\mathrm{f}^4}{4},
\end{equation}
where $R_\mathrm{f}$ is the radius of the cross section of the filament.
For crosslinked filaments, if we ignore the contribution of the crosslinkers, we can calculate their second moment of area.
To do so, it is convenient to apply the parallel axis theorem, which allows the second moment of area to be calculated when an expression is known for a given cross-sectional shape, but with the bending axis having been translated relative to it.
This gives
\begin{equation}
    I = \mleft( I_\mathrm{cm} + A d^2 \mright),
\end{equation}
where $I_\mathrm{cm}$ is the known second moment of area when the center of geometry is on the bending axis, and $d$ is the distance from the bending axis (\cref{fig:crosssection}).
For a given configuration of filaments, we then have
\begin{equation}
    I_i = \sum_j \pi R_\mathrm{f}^2 \mleft( \frac{R_\mathrm{f}^2}{4} + d_j^2 \mright),
\end{equation}
where $j$ is the filament index.

Because of the way anillin bundles actin into sheets, for anillin-actin rings, the bending resistance is minimized when $d = 0$ for all filaments (\cref{fig:crosssection}(b)).
In other words, such a ring will have its minimum free energy when the filaments form a sheet-like configuration such that each filament is bent to the same extent (e.g.~\namecref{fig:crosssection} \ref{fig:diagram}(b) of the main text and \cref{fig:crosssection}(b)), in contrast to a configuration where the filaments are stacked such that they have different curvatures (\cref{fig:crosssection}(a)).
The second moment of area of a segment is then simply given by the number of filaments in that segment multiplied by the second moment of area of a single filament.
This allows the bending energy of the whole ring to be calculated by summing over each filament instead of over the segments, giving
\begin{equation}
    U_\mathrm{B} = \frac{\Nf E I \Lf}{2 R^2}. \label{eq:bending-energy}
\end{equation}

\begin{figure}
    \centering
    \includegraphics{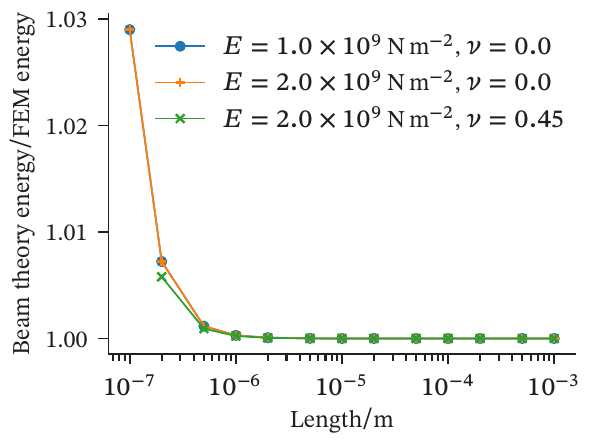}
    \caption{
        Ratio of beam-theory energy to geometrically-nonlinear-elasticity energy of a single filament, face-to-face ring as a function of filament length.
        The geometrically-nonlinear energies are calculated by first solving for the deformation, given the constraints that the end faces of the filament must be touching.
        The finite element method was used to solve these equations, in particular the deal.II library \cite{arndt2021_sup,arndt2021b_sup}.
        A rectangular-cuboid geometry was used for simplicity, with the height taken to be twice the radius used by \citet{kojima1994_sup} (\SI{2.8}{\nano\meter}) to calculate the Young's modulus.
        To ensure good convergence, 10 000 divisions were made along the filament axis, while no divisions were made in perpendicular axes.
        \label{fig:beam-vs-fem}
    }
\end{figure}

We test whether the assumptions of beam theory in this regime are valid by comparing the energy calculated from \cref{eq:bending-energy} to the energy of the rings calculated with geometrically nonlinear elasticity, or the Saint Venant-Kirchhoff model.
This model of elasticity is still linear in the stress-strain relation, but includes nonlinearity in the strain tensor.
In \cref{fig:beam-vs-fem} we plot the ratio of the beam-theory energy to the geometrically-nonlinear-elasticity energy for a single filament forming a ring in which the faces of the two ends of the filament are constrained to touch for a range of filament lengths.
At short enough filament lengths, regardless of the bending rigidity, the two models diverge, but within the range of filament lengths considered here, the error involved with the simpler beam theory model is insignificant.
Geometrically nonlinear elasticity has an additional parameter, the Poisson's ratio $\nu$, and while it can affect the regime of validity of beam theory, it also has no significant effect on the results in the range of filament lengths relevant here.
The validity of beam theory in this regime is also supported by experimental results~\cite{arai1999_sup}.

\section{Ring-constriction dynamics}

We assume the system is in the overdamped limit, in which case the Langevin equation is a first-order stochastic differential equation.
To fully specify the dynamics, we must also specify the interpretation used for the stochastic integral~\cite{sokolov2010_sup}.
Here, we use the Ito interpretation, which, when the friction is position dependent, results in a spurious drift term~\cite{yang2013_sup}, giving
\begin{equation}
    \mathrm{d} R = \mleft[ \frac{\ForceR}{\zetaR} + \deriv{}{R} \mleft( \frac{\kb T}{\zetaR} \mright) \mright] \mathrm{dt} + \sqrt{\frac{2 \kb T}{\zetaR}} \mathrm{d} W, \label{eq:dynamics}
\end{equation}
where $\ForceR$ is the constriction force, $\zetaR$ is the friction coefficient for ring constriction, $\kb$ is Boltzmann's constant, $T$ is the temperature, and $\mathrm{d} W$ is a Wiener increment.
The constriction force referred to in the main text is defined as
\begin{equation}
    F = \ForceR = -\deriv{U}{R},
\end{equation}
where $U$ is the ring energy discussed in the main text.
An expression for this force is given in \namecref{eq:dynamics} \ref{eq:tracks-force} of the main text, so to solve for the dynamics, the only additional expression needed is for the friction coefficient.

\section*{Derivation of friction coefficient}

In this section, we derive the friction coefficient for ring-constriction dynamics.
In particular, we show how the friction coefficient for ring constriction can be expressed in terms of the friction coefficient of a single overlap and the two parameters that characterize the ring topology, $\Nf$ and $\Nsca$.
We then derive an exact form for the friction coefficient of a single overlap with a constant number of bound crosslinkers, and an approximate form that agrees well with the exact one.
Finally, we use the result in the regime with a constant number of crosslinkers to derive the friction coefficient in the regime with a constant crosslinker concentration.

\subsection{Relation between ring friction and single-overlap friction}

The friction coefficient for radial ring-constriction dynamics $\zetaR$ can be related to that of a single overlap $\zeta$.
To do so, first consider the balance of forces in a slice of the ring.
The friction force is determined by the friction coefficients of individual overlaps, the rate at which the filaments slide, and their connectivity in the ring.
The constriction rate of individual overlaps connected in series, relative to the constriction rate of the circumference $C$, decreases with the number of overlaps connected in series, which is $\Nsca$; the friction associated with the circumference is thus decreased by a factor of $\nicefrac{1}{\Nsca}$.
On the other hand, the friction of overlaps acting in parallel must be summed to calculate the total friction.
With the assumptions of the main text, where the total number of overlaps $\No = 2 \Nf - \Nsca$, there are $\nicefrac{\No}{\Nsca}$ overlaps in parallel in a given slice of a ring; the friction is thus increased by a factor of $\nicefrac{\No}{\Nsca}$.
Taking into account both effects leads to a total friction given by
\begin{subequations}
\begin{align}
    F_\mathrm{f} &= \frac{\No}{\Nsca^2} \zeta \deriv{C}{t}\\
                 &= \frac{\No}{\Nsca^2} \zeta \deriv{C}{R} \deriv{R}{t}\\
                 &= \frac{2 \pi \No}{\Nsca^2} \zeta \deriv{R}{t}.
\end{align}
\end{subequations}
This friction force balances with the circumferential constriction force, $F_\mathrm{C}$,
\begin{subequations}
\begin{align}
    F_\mathrm{C} &= F_\mathrm{f}\\
    -\deriv{U}{R} \deriv{R}{C} &= \frac{2 \pi \No}{\Nsca^2} \zeta \deriv{R}{t},
\end{align}
\end{subequations}
which, upon rearranging, gives
\begin{subequations}
\begin{align}
    -\deriv{U}{R} &= \frac{4 \pi^2 \No}{\Nsca^2} \zeta \deriv{R}{t}\\
    \ForceR &= \zetaR \deriv{R}{t},
\end{align}
\end{subequations}
thus giving the desired relation between the friction coefficient for radial ring-constriction dynamics and the friction coefficient of a single overlap,
\begin{equation}
    \zetaR = \frac{4 \pi^2 \No}{\Nsca^2} \zeta. \label{eq:zetaR}
\end{equation}
We give an alternative derivation of the relation between $\zetaR$ and $\zeta$ in the Appendix.


\begin{figure}
    \centering
    \includegraphics{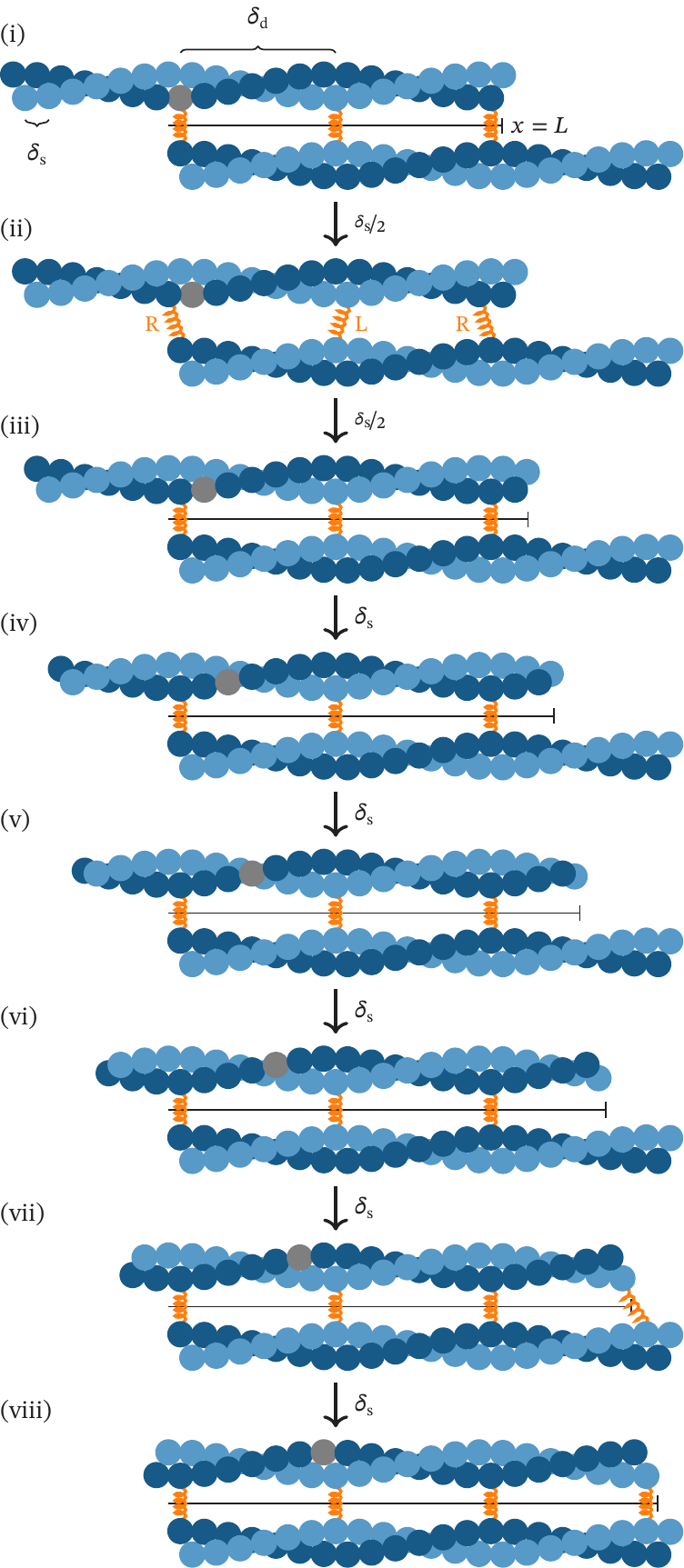}
    \caption{
        Diagrams of a mobile filament sliding over a fixed filament to increase the overlap $L$ by $\deltad$ from (i) to (viii).
        One monomer is coloured in grey to better illustrate the twisting of the top filament.
        Over the first two increments, the top filament is shown to slide in increments of $\nicefrac{\deltas}{2}$ to illustrate in (ii) an example of a configuration at the peak of the barrier to sliding; $R$ denotes a crosslinker that pulls the mobile top filament to the right, while $L$ denotes a crosslinker that pulls the mobile top filament to the left.
        All other configurations are shown when the binding sites are in register.
        In every configuration, all available crosslinker sites are filled; i.e.~the configurations are shown at maximum occupancy.
        In (vii), the number of sites that can be crosslinked increases from three to four, as at this point the binding site on the far right of the mobile filament is within a distance of $\deltas$ from a nearly aligned binding site on the fixed filament.
        \label{fig:sliding}
    }
\end{figure}

\subsection{Derivation of single-overlap friction coefficient}

Wierenga and Ten Wolde~\cite{wierenga2020_sup} and \citet{wierenga2021_sup} have developed expressions for the friction coefficient of a single overlap $\zeta$.
Their approach is to relate the height of the free-energy barrier $\upDelta \mathcal{F}^\ddag$ for filament sliding to the friction coefficient via two expressions.
First, for a system with discrete jumps on a lattice with spacing $\delta$, the friction coefficient can be related to the jump rate in one direction $r$ via the Einstein equation,
\begin{equation}
    \zeta = \frac{\kb T}{D} = \frac{\kb T}{\delta^2 r}. \label{eq:einstein}
\end{equation}
Then, the jump rate can be related to the free-energy barrier with the Arrhenius equation,
\begin{equation}
    r = r_0 \expsup{-\beta \upDelta \mathcal{F}^\ddag}, \label{eq:arrhenius}
\end{equation}
where $\beta^{-1} = \kb T$, and $r_0$ is the jump-rate prefactor.

In the Ase1-microtubule system, the microtubules are crosslinked at discrete sites, where the lattice spacing $\delta$ is set by the tubulin-dimer distance.
Microtubules slide in increments of $\delta$ by a collective rearrangement, wherein all crosslinkers hop to a neighbouring site~\cite{wierenga2020_sup}.
In the process of hopping, the crosslinkers are stretched from their equilibrium length.

In actin, whose filaments form a double helix, the filaments are also crosslinked at discrete sites, but the crosslinker spacing is set by the helical pitch rather than by the monomer-monomer distance $\deltas$, as the sites only line up every half-turn of the filaments (\cref{fig:sliding}).
This distance $\deltad$ is far larger than the size of a crosslinker, making the mechanism observed in microtubule sliding seem unfeasible.
However, studies of myosin-driven actin sliding show that actin filaments tend to twirl, or rotate, about their axes as they are slid by myosin motors anchored to a surface~\cite{jegou2020_sup,ropars2016_sup,vilfan2009_sup,beausang2008_sup,ali2002_sup,sase1997_sup}.
We therefore argue that the same collective hopping can occur, but with the actin filament rotating while sliding.
Then, each collective rearrangement would occur with the monomer-monomer distance of one of the two intertwined filaments in the double helix, thus with spacing $\deltas$ instead of $\deltad$ (\cref{fig:sliding}).
If we assume that crosslinkers can stretch to bind to a site a distance no more than $\deltas$ away, then the number of overlapping sites $\ell$ can be related to the overlap length $L$ with
\begin{subequations}
\begin{align}
    \ell &= \mleft\lfloor \frac{1}{\deltad} (L + \deltas) \mright\rfloor + 1 \label{eq:exactl}\\
         &\approx \mleft\lfloor \frac{L}{\deltad} \mright\rfloor + 1 \label{eq:discretel}\\
         &\approx \frac{L}{\deltad} + 1,
\end{align}
\end{subequations}
where first two lines use the floor function (see \cref{fig:sliding}(vii) to understand \cref{eq:exactl}), and the third line is the continuous approximation of $\ell$, which we will use henceforth unless otherwise stated (or implied by use in a combinatorial expression).

\subsubsection{Friction coefficient for constant number of crosslinkers}

Wierenga and Ten Wolde~\cite{wierenga2020_sup} derive an exact form for the free-energy barrier of the sliding of an overlapping filament, in their case a microtubule, when the number of bound crosslinkers is constant.
We follow the same steps to derive an exact form for the free-energy barrier to actin-filament sliding, which involve writing out the potential, determining the number of microstates, and selecting a reaction coordinate for filament sliding.
In their derivation, they consider one of the filaments to be fixed, and then partition the crosslinkers into those that pull the mobile filament to the right, and those that pull it to the left (\cref{fig:sliding}(ii)).
If we consider a fixed bottom filament and a mobile top filament, and define $x$ as the modulated position of the filament tip with $x = L \bmod \deltas$, then the potential energy can be written as
\begin{subequations}
\begin{align}
    U(x, \NR) &= \frac{1}{2} k x^2 (\Nd - \NR) + \frac{1}{2} k (\deltas - x)^2 \NR\\
      &= \frac{1}{2} k \deltas^2 \Nd \mleft[ \mleft( \frac{x}{\deltas} - \frac{\NR}{\Nd} \mright)^2 + \frac{\NR}{\Nd} \mleft( 1 - \frac{\NR}{\Nd} \mright) \mright],
\end{align}
\end{subequations}
where $\Nd$ is the number of bound crosslinkers, $\NR$ is the number of right-pulling crosslinkers (\cref{fig:sliding}(ii)), and $k$ is the spring constant of the crosslinker.
In their system, crosslinkers can be bound at adjacent sites, which complicates the calculation of the entropic contribution to the free energy, as the right-pulling crosslinkers can block binding of left-pulling crosslinkers to adjacent sites.
Here, because the crosslinker-crosslinker spacing $\deltad$ is much larger than the crosslinker hopping distance $\deltas$ (\cref{fig:sliding}(i)), they cannot block each other when stretched, simplifying the calculation of the entropic contribution.
Then, the number of microstates as a function of the number of right-pulling crosslinkers is
\begin{equation}
    \Omega(\NR) = \binom{\ell}{\Nd} \binom{\Nd}{\NR}.
\end{equation}
Wierenga and Ten Wolde~\cite{wierenga2020_sup} found a good description of the reaction coordinate for sliding a distance of $\deltas$,
\begin{equation}
    \alpha = \frac{1}{2} \mleft( \frac{x}{\deltas} + \frac{\NR}{\Nd} \mright).
\end{equation}
Using this reaction coordinate, the partition function is then
\begin{multline}
    Q(\alpha, \ell) = \sum_{\NR = 0}^{\Nd} \mathbb{1} \mleft( 0 \leq 2 \alpha - \frac{\NR}{\Nd} \leq 1 \mright) \binom{\ell}{\Nd} \binom{\Nd}{\NR}\\
    \times \expsup{-\frac{k \deltas^2 \Nd}{2 \kb T} \mleft[ 4 \mleft( \alpha - \frac{\NR}{\Nd} \mright)^2 + \frac{\NR}{\Nd} \mleft( 1 - \frac{\NR}{\Nd} \mright) \mright]},
\end{multline}
where $\mathbb{1}(a)$ is the indicator function, being 1 if $a$ is true, and 0 otherwise.
Because the number of binding sites only changes at intervals of $\deltad$, while sliding occurs on the scale of $\deltas$, $\ell$ will not change for most barrier crossings (\cref{fig:sliding}).
With constant $\ell$, the barrier to sliding a distance of $\deltas$ can be calculated with the ratio of the partition functions at $\alpha = 0$,
\begin{equation}
    Q(\alpha = 0, \ell) = \binom{\ell}{\Nd}, \label{eq:entpartregister}
\end{equation}
and $\alpha = \nicefrac{1}{2}$,
\begin{equation}
    Q(\alpha = \nicefrac{1}{2}, \ell) = \sum_{\NR = 0}^{\Nd} \binom{\ell}{\Nd} \binom{\Nd}{\NR} \expsup{-\frac{k \deltas^2 \Nd}{2 \kb T} \mleft[ \frac{3}{\Nd^2} \mleft( \NR - \frac{\Nd}{2} \mright)^2 + \frac{1}{4} \mright]},
\end{equation}
to give
\begin{subequations}
\begin{align}
    \beta \upDelta \mathcal{F}^\ddag &= -\ln \mleft[ \frac{Q ( \alpha = \nicefrac{1}{2}, \ell )}{Q ( \alpha = 0, \ell)} \mright]\\
                                     &= \ln \mleft[ \sum_{\NR = 0}^{\Nd} \binom{\Nd}{\NR}^{-1} \expsup{\frac{k \deltas^2 \Nd}{2 \kb T} \mleft[ \frac{3}{\Nd^2} \mleft( \NR - \frac{\Nd}{2} \mright)^2 + \frac{1}{4} \mright]} \mright]. \label{eq:barrierexact}
\end{align}
\end{subequations}
When sliding also increases the number of sites, the barrier height will change, and it will also become dependent on whether the overlap is increasing or decreasing; as a first approximation, we will ignore this effect.

To derive an approximate exponential form for the barrier, we can again follow Wierenga and Ten Wolde~\cite{wierenga2020_sup}.
In their derivation, they use a Gaussian approximation for the binomials, which involves expanding at $\NR = \nicefrac{\Nd}{2}$, which is where the summand in \cref{eq:barrierexact} peaks,
\begin{equation}
    \binom{\Nd}{\NR} \approx \sqrt{\frac{2}{\pi \Nd}} \expsup{\Nd \ln 2 -\frac{2}{\Nd} \mleft(\NR - \frac{\Nd}{2} \mright)^2}.
\end{equation}
Using this approximation, the partition function ratio here is then
\begin{multline}
    \frac{Q ( \alpha = \nicefrac{1}{2}, \ell )}{Q ( \alpha = 0, \ell)} \approx \sqrt{\frac{2}{\pi \Nd}} \expsup{\Nd \ln 2 -\frac{k \deltas^2 \Nd}{8 \kb T}}\\
    \times \sum_{\NR = 0}^{\Nd} \expsup{-\mleft( \frac{3k \deltas^2}{2 \Nd \kb T} \frac{2}{\Nd} \mright) \mleft( \NR - \frac{\Nd}{2} \mright)^2}.
\end{multline}
With the assumption that $\Nd$ is large enough to approximate as a continuous variable, the sum in the above expression can be approximated as an integral over the entire real line, which, with the appropriate substitution, results in a Gaussian integral.
Solving the integral, rearranging, and plugging the approximation of the partition function ratio back into the expression for the barrier gives
\begin{equation}
    \beta \upDelta \mathcal{F}^\ddag \approx A + B \Nd, \label{eq:barrierexp}
\end{equation}
where
\begin{equation}
    A = \frac{1}{2} \ln \mleft( 1 + \frac{3 k \deltas^2}{4 \kb T} \mright),
\end{equation}
and
\begin{equation}
    B = \frac{\deltas^2 k}{8 \kb T} - \ln 2. \label{eq:B}
\end{equation}
The resulting friction coefficient is
\begin{equation}
    \zeta = \zeta_0 \expsup{B \Nd}, \label{eq:zetaexp}
\end{equation}
where
\begin{equation}
    \zeta_0 = \frac{\kb T}{\deltas^2 r_0} \sqrt{1 + \frac{3 k \deltas^2}{4 \kb T}}. \label{eq:zeta0}
\end{equation}
Here, unlike the form derived by Wierenga and Ten Wolde~\cite{wierenga2020_sup}, where the crosslinkers can block each other, the friction coefficient does not have a superexponential dependence on the density.
The friction depends only exponentially on the number of bound crosslinker $\Nd$, reflecting the fact that the filament hopping rate depends exponentially on the barrier to hopping, with the latter scaling linearly with $\Nd$.

\begin{figure}
    \centering
    \includegraphics{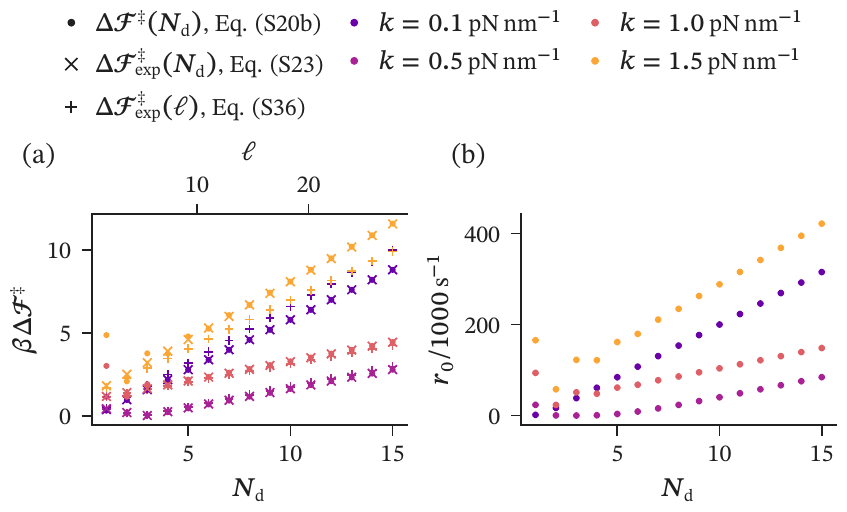}
    \caption{
        \label{fig:barriers-r0}
        (a) Comparison of the free-energy barrier as a function of $\Nd$ calculated with the exact expression (\cref{eq:barrierexact}, ``$\upDelta \mathcal{F}^\ddag(\Nd)$''), and with the exponential approximation (\cref{eq:barrierexp}, ``$\upDelta \mathcal{F}^\ddag_\mathrm{exp}(\Nd)$'') and as a function of $\ell$ with the constant $\ce{[X]}$ expression (\cref{eq:barriercx}, ``$\upDelta \mathcal{F}^\ddag_\mathrm{exp}(\ell)$''), each with four different values of $k$ that encompass the range of measured values of $k$ for crosslinkers~\cite{duke1999_sup,wierenga2020_sup,jeney2004_sup,ahmadzadeh2014_sup}.
        The exponential approximation with constant $\Nd$ agrees very well with the exact expression for $\Nd \ge 4$; for $\Nd = 2$ and $\Nd = 3$, the exponential form gives a slightly higher and lower barrier, respectively, while  $\Nd = 1$, the exponential form gives a significantly lower barrier.
        The values of $\ell$ are aligned with the values of $\Nd$ that equal the equilibrium average (see \cref{eq:occupancy} for the average occupancy).
        The constant $\ce{[X]}$ expression agrees well with the constant $\Nd$ expressions, with only small deviations at values of $k$ and $\Nd$ that lead to large barriers.
        For a given $\Nd$, there is an optimal value of $k$ that minimizes the barrier.
        (b) Plot of the hopping rate prefactor $r_0$ as a function of $\Nd$ for four different values of $k$.
        The dependence of $r_0$ is roughly linear, in contrast to the exponential dependence in the microtubule system of Ref.~\cite{wierenga2020_sup}.
        Data for both plots were produced with the parameters given in \cref{tab:params}.
    }
\end{figure}

With the parameters in \cref{tab:params}, the barrier calculated with the exponential approximation agrees well with the barrier calculated with the exact form if $\Nd \ge 4$ (\cref{fig:barriers-r0}(a)).
With both calculation methods, the barrier does not simply decrease with decreasing $k$, but increases when $k$ is sufficiently low.
In the exponential approximation to the barrier, \cref{eq:barrierexp}, the relationship is clear: whether the barrier increases or decreases with $k$ for a given $\Nd$ depends on the sign of $B$.
The sign determines whether the energy of stretching these linkers (the first term of the right-hand side (RHS) of \cref{eq:B}) or the entropy of mixing the left- and right-pulling crosslinkers (the second term of the RHS of \cref{eq:B}) dominates.
When $B < 0$ and $B \Nd > -A$, the peak of the barrier is shifted to $\alpha = 0$, and the trough to $\alpha = \nicefrac{1}{2}$.
This implies that for a given $\Nd$, there is a value of $k$ that will minimize the friction, specifically that which leads to $A = -B \Nd$.

\subsubsection{Friction coefficient for constant crosslinker concentration}

\citet{wierenga2021_sup} uses the approximate form for the free-energy barrier to sliding with constant $\Nd$ to derive the friction coefficient in the regime with fast crosslinker binding, where the crosslinker concentration is held constant instead.
As in the previous section, we follow their approach to derive the friction coefficient for actin-filament sliding.
The partition function is now the weighted sum of partition functions with constant $\Nd$.
The weights include both the usual grand-ensemble weight and the weight associated with the free energy of binding.
For bound crosslinkers, the weight is
\begin{equation}
    \xi_\mathrm{d} = \mathrm{e}^{\beta \mu_\mathrm{X}} \mathrm{e}^{-\beta \upDelta \mathcal{F}_\mathrm{d}},
\end{equation}
where $\mu_\mathrm{X}$ is the chemical potential of the crosslinkers in solution, and $\upDelta \mathcal{F}_\mathrm{d}$ is the free energy of binding a crosslinker from solution.
If we assume the crosslinkers act ideally in solution and use the definition of the dissociation constants in terms of the free energy, then we can simplify the weight to
\begin{equation}
    \xi_\mathrm{d} = \frac{\ce{[X]}}{\Kd}, \label{eq:doubleweight}
\end{equation}
where $\Kd$ is the dissociation constant for crosslinkers binding to two filaments from solution.
The grand partition function also includes the effect of crosslinkers bound with only one end, or in other words, to only a single filament, on the average number of bound crosslinkers.
We will refer to these as ``singly bound'' sites or crosslinkers.
The weight for singly bound crosslinkers can similarly be written as
\begin{equation}
    \xi_\mathrm{s} = \frac{\ce{[X]}}{\Ks}, \label{eq:singleweight}
\end{equation}
where $\Ks$ is the dissociation constant for crosslinkers binding to a single filament from solution.
Using these weights, the grand partition function is
\begin{subequations}
\begin{align}
    \Xi(\alpha, \ell) &= \sum_{\Nd = 0}^{\ell} \xi_\mathrm{d}^{\Nd} Q(\alpha, \ell) \sum_{\Ns = 0}^{\ell_\mathrm{f} - 2 \Nd} \binom{2(\ell_\mathrm{f} - \Nd)}{\Ns} \xi_\mathrm{s}^{\Ns}\\
                      &= \sum_{\Nd = 0}^{\ell} \xi_\mathrm{d}^{\Nd} Q(\alpha, \ell) (1 + \xi_\mathrm{s})^{2(\ell_\mathrm{f} - \Nd)}\\
                      &= (1 + \xi_\mathrm{s})^{2 \ell_\mathrm{f}} \sum_{\Nd = 0}^{\ell} \xi^{\Nd} Q(\alpha, \ell),
\end{align}
\end{subequations}
where $\Ns$ is the number of singly bound crosslinkers, $\ell_\mathrm{f}$ is the number of binding sites on a single filament, and
\begin{equation}
    \xi = \frac{\xi_\mathrm{d}}{(1 + \xi_\mathrm{s})^2}.
\end{equation}
If we redefine the reference state, we can drop the term before the sum in the last line to give
\begin{equation}
    \Xi(\alpha, \ell) = \sum_{\Nd = 0}^{\ell} \xi^{\Nd} Q(\alpha, \ell). \label{eq:condpart}
\end{equation}
For $\alpha = 0$,
\begin{subequations}
\begin{align}
    \Xi(\alpha = 0, \ell) &= \sum_{\Nd = 0}^{\ell} \xi^{\Nd} Q(\alpha = 0, \ell)\\
    &= \sum_{\Nd = 0}^{\ell} \xi^{\Nd} \binom{\ell}{\Nd}\\
    &= (1 + \xi)^\ell, \label{eq:condpartregister}
\end{align}
\end{subequations}
while for $\alpha = \nicefrac{1}{2}$,
\begin{subequations}
\begin{align}
    \Xi(\alpha = \nicefrac{1}{2}, \ell) &= \sum_{\Nd = 0}^{\ell} \xi^{\Nd} Q(\alpha = \nicefrac{1}{2}, \ell)\\
    &= \sum_{\Nd = 0}^{\ell} \xi^{\Nd} \expsup{-\beta \upDelta \mathcal{F}^\ddag} Q(\alpha = 0, \ell)\\
    &\approx \sum_{\Nd = 0}^{\ell} \xi^{\Nd} \expsup{A + B \Nd} \binom{\ell}{\Nd}\\
    &= \expsup{-A} \mleft( 1 + \xi \expsup{-B} \mright)^\ell.
\end{align}
\end{subequations}
Taken together, the barrier becomes
\begin{equation}
    \beta \upDelta \mathcal{F}^\ddag \approx A + \ell \ln \mleft[ \frac{1 + \xi}{1 + \xi \expsup{-B}} \mright], \label{eq:barriercx}
\end{equation}
which gives a friction coefficient of
\begin{equation}
    \zeta = \zeta_0 \mleft[ \frac{1 + \xi}{1 + \xi \expsup{-B}} \mright]^{\mleft( \frac{L}{\deltad} + 1 \mright)} \label{eq:zetacx}.
\end{equation}

We can compare the barrier in this regime to one with constant $\Nd$ if we set $\Nd$ to its equilibrium value for a given $\ell$.
If we only consider states where the binding sites are in register, around which the system will spend the majority of the time, then the equilibrium occupancy of a single site is
\begin{subequations}
\begin{align}
    \mleft\langle \frac{\Nd}{\ell} \mright\rangle &= \frac{\xi_\mathrm{d}}{(1 + \xi_\mathrm{s})^2 + \xi_\mathrm{d}}\\
                                      &= 1 + \frac{{\Ks}^2 \ce{[X]}}{\Kd ( \Ks + \ce{[X]})^2}. \label{eq:occupancy}
\end{align}
\end{subequations}
Because the sites are independent, $\langle \Nd \rangle = \ell \langle \nicefrac{\Nd}{\ell} \rangle$.
From \cref{fig:barriers-r0}(a), we see good agreement with the expressions for constant $\Nd$, with noticeable deviation only at values of $k$ and $\Nd$ that lead to large barriers.

\subsubsection{Estimation of jump-rate prefactor}

\citet{wierenga2021_sup} uses Kramers' theory to derive an approximate expression for the prefactor that appears in the Arrhenius equation (\cref{eq:arrhenius}), $r_0$.
Kramers' theory gives $r_0$ in terms of the free-energy profile $\upDelta \mathcal{F}(\alpha)$ and the diffusion coefficient along the barrier $D(\alpha)$ as functions of the reaction coordinate $\alpha$; with some approximations, this produces an expression for $r_0$ in terms of the free-energy barrier and the diffusion coefficient of the filament at the peak of the barrier, $D^\ddag$,
\begin{subequations}
\begin{align}
    r_0 &= \mleft[ \int_0^1 \frac{\expsup{\beta[\upDelta \mathcal{F}(\alpha) - \upDelta \mathcal{F}^\ddag]}}{D(\alpha)} \mathrm{d} \alpha \int_0^1 \expsup{-\beta \upDelta \mathcal{F}(\alpha)} \mathrm{d} \alpha \mright]^{-1} \label{eq:r0exact}\\
        &\approx \frac{8 (\beta \upDelta \mathcal{F}^\ddag)^3 D^\ddag \pi}{8 (\beta \upDelta \mathcal{F}^\ddag)^2 + 4 \beta \upDelta \mathcal{F}^\ddag + 5}. \label{eq:r0}
\end{align}
\end{subequations}
To derive this expression, they approximate the free-energy profile with a sinusoidal function, using the barrier peak to set the amplitude, and use an asymptotic expansion valid for sufficiently larger barriers.
Further, they assume that $D(\alpha)$ can be approximated by $D^\ddag$, which they justify by the fact that the contribution to the integral in \cref{eq:r0exact} is largest around the barrier peak and that the variation in the diffusion coefficient with respect to $\alpha$ is small relative to the variation in the free-energy profile.

We can use \cref{eq:r0} to estimate the prefactor for actin-filament sliding, but the expression for $D^\ddag$, in addition to the expression for $\upDelta \mathcal{F}^\ddag$, will differ from those of microtubules.
For the free-energy barrier, we use the exact expression given in \cref{eq:barrierexact}.
The diffusion coefficient at the peak is approximated in terms of the diffusion coefficient of a filament in water, $D_\mathrm{m}$, the hopping rate of a singly bound crosslinker, $h_0$, and the number of bound crosslinkers, $\Nd$,
\begin{equation}
    D^\ddag = \frac{1}{4} \mleft( \frac{D_\mathrm{m}}{\deltas^2} + \frac{h_0}{\Nd} \mright). \label{eq:Dbarrier}
\end{equation}
The expression here differs from that in Ref.~\cite{wierenga2021_sup} because, as discussed above, in the anillin-actin system, bound crosslinkers cannot block each other from hopping, so the total hopping rate does not decrease with a higher density of bound crosslinkers.
The diffusion coefficient of an actin filament can be approximated with an expression derived for rigid rods for the diffusion coefficient of translation parallel to the cylindrical axis~\cite{li2004_sup,broersma1981_sup,broersma1960_sup},
\begin{equation}
    D_\mathrm{m} \approx D_\| = \frac{\kb T \mleft(\ln \frac{\Lf}{\Df} - \gamma_\| \mright)}{2 \pi \eta \Lf}, \label{eq:Dm}
\end{equation}
where $\Df$ is the diameter of an actin filament, $\gamma_\|$ is an end-correction coefficient, and $\eta$ is the viscosity of the fluid.
Here we use the value for $\gamma_\|$ used by \citet{li2004_sup}, while for $\eta$ we use the value of pure water at \SI{298}{\kelvin}~\cite{crc97_sup}.
The remaining term in $D^\ddag$, $h_0$, can be estimated from the diffusion constant of singly bound anillin, which has been experimentally measured~\cite{kucera2021_sup}, with
\begin{equation}
    h_0 = \frac{D_\mathrm{s}}{\deltas^2}.
\end{equation}
However, $D_\mathrm{m}$ is two orders of magnitude larger than $D_\mathrm{s}$, so the contribution of the $h_0$ term in \cref{eq:Dbarrier} is insignificant.

Because the hopping rate depends exponentially on $\Nd$ through the barrier but only linearly through $r_0$, we calculate one value of $r_0$, at $\Nd = 4$, for a given value of the spring constant $k$.
There is also some evidence to expect that $r_0$ should be approximately constant with respect to changes in $\Nd$.
To test \cref{eq:r0}, \citet{wierenga2021_sup} compared values for a range of $\Nd$ at several overlap lengths to Brownian dynamics simulations that explicitly modeled the collective hopping of crosslinkers to allow sliding of the filament.
Overall, their expression was shown to give values of $r_0$ with the right order of magnitude, but with an exponential dependence on the crosslinker occupancy.
The simulation values of $r_0$ tended to be relatively flat with changes in the occupancy and the total number of bound crosslinkers.
Our expressions for $\upDelta \mathcal{F}^\ddag$ and $D^\ddag$ are linearly dependent on the total number of bound crosslinkers, but not exponentially on the density.
A plot of $r_0$ shows that it increases roughly linearly with $\Nd$ (\cref{fig:barriers-r0}(b)).
While we do not have simulations with explicit crosslinker hopping for the anillin-actin system to compare the analytical approximation to, it should be similar to the microtubule system at low densities.


\section*{Discussion of model parameters}

\begin{table*}[ht]
    \centering
    \begin{tabular}{l p{12cm} c l}
        \toprule
        Parameter & Description & Value & Source\\
        \midrule
        $\Ks$ & Dissociation constant for crosslinker binding from solution to a single filament & \SI{7.82}{\nano\molar} & \cite{kucera2021_sup} \\
        $\Kd$ & Dissociation constant for crosslinker binding from solution to two filaments & $\Ks / 5$ & \\
        $\ce[X]$ & Crosslinker concentration & \SI{12}{\nano\molar} & \cite{kucera2021_sup} \\
        $\deltad$ & Spacing between crosslinker binding sites in overlapping actin filaments & \SI{36}{\nano\meter} & \cite{dominguez2011_sup} \\
        $\deltas$ & Spacing between actin monomers & $\frac{2 \deltad}{13} \approx \SI{6}{\nano\meter}$ & \cite{dominguez2011_sup} \\
        $T$ & Temperature & 300 & \cite{kucera2021_sup} \\
        $\Nsca$ & Number of scaffold ring filaments & \numrange{2}{8} & \\
        $\Nf$ & Total number of force-generating filaments & \numrange{2}{8} & \\
        $\No$ & Number of overlaps & $2 \Nf - \Nsca$ & \\
        $EI$ & Bending rigidity of an actin filament & \SI{3.6e26}{\newton\meter\squared} & \cite{isambert1995_sup} \\
        $\Lf$ & Filament length & \SIrange[range-units=single]{1}{6}{\micro\meter} & \\
        $k$ & Spring constant of a crosslinker & \SIrange[range-units=single]{1.0}{2.0}{\pico\newton\per\nano\meter} & \\
        $D_\mathrm{s}$ & Diffusion coefficient of singly bound anillin & \SI{8.8e-3}{\micro\meter\squared\per\second} & \cite{kucera2021_sup} \\
        $\eta$ & Viscosity of water & \SI{8.9e-4}{\newton\second\per\meter\squared} & \cite{crc97_sup} \\
        $\gamma_\|$ & End-correction coefficient for the parallel translation diffusion coefficient of a rod & $-0.114$ & \cite{li2004_sup,broersma1981_sup} \\
        $\Df$ & Diameter of an actin filament & \SI{8}{\nano\meter}& \cite{li2004_sup} \\
        \bottomrule
    \end{tabular}
    \caption{
        \label{tab:params}
        Parameters used in calculations.
    }
\end{table*}

The model has four constants which must be known to calculate forces: the bending rigidity $EI$ of actin, the spacing between anillin binding sites on actin $\deltad$, the dissociation constant of anillin binding to a single actin filament $\Ks$, and the dissociation constant of anillin crosslinking two actin filaments $\Kd$.
To model the dynamics, an additional three parameters are needed: the actin monomer-monomer distance $\deltas$, the spring constant of a crosslinker $k$, and the jump-rate prefactor for crosslinker hopping $r_0$.
Because we use an analytical approximation to estimate $r_0$, that single parameter becomes four: the diffusion coefficient of a singly bound anillin $D_\mathrm{s}$, the viscosity of water $\eta$, the end-correction coefficient of the parallel translation diffusion coefficient of a rod $\gamma_\|$, and the diameter of an actin filament $\Df$.
The model parameters are collected in \cref{tab:params}.

The bending rigidity of single actin filaments has been measured and calculated in many studies~\cite{mameren2009_sup,brangwynne2007_sup,arai1999_sup,isambert1995_sup,kojima1994_sup,ott1993_sup}.
Van Mameren \textit{et al.}~\cite{mameren2009_sup} determined a value of \SI{7.1e-26}{\newton\meter^2} by bending filaments, measuring the response, and fitting a nonlinear elastic model.
Their result is broadly similar to what has been found in older studies that used other methods that actively perturb filaments~\cite{arai1999_sup,kojima1994_sup}, as well as to results with methods that passively observe thermal fluctuations~\cite{brangwynne2007_sup,isambert1995_sup,ott1993_sup,gittes1993_sup}.
However, most of these studies considered actin filaments stabilized by the presence of phalloidin.
In the ring experiments of \citet{kucera2021_sup}, they use conditions both with and without phalloidin.
\citet{isambert1995_sup} found that without the presence of phalloidin, actin filaments are about twice as flexible.
This corresponds to a bending rigidity of \SI{3.6e-26}{\newton\meter^2}.

One way to estimate $\deltad$ would be to consider that, on one side of the helical actin filaments, a monomer repeats about about every \SI{6}{\nano\meter}~\cite{dominguez2011_sup}.
However, \citet{matsuda2020_sup} found that anillin only binds when the twist of the helix of the two filaments is properly aligned, which only occurs every \SI{36}{\nano\meter}~\cite{dominguez2011_sup}, and so we choose this value for $\deltad$.
Since there are 13 monomers along one strand every $2 \deltad$, we define $\deltas = \nicefrac{2 \deltad}{13} \approx \SI{6}{\nano\meter}$.

Three studies have measured a dissociation constant for anillin with actin.
\citet{matsuda2020_sup} reported a value of \SI{1.2}{\micro\molar}, while \citet{jananji2017_sup} reported an even weaker value of \SI{4.3}{\micro\molar}.
For comparison, in the original study of force generation via passive crosslinkers, Ase1 was found to bind to microtubules with dissociation constants three orders of magnitude smaller, in the \si{\nano\molar} range.
In both of the experiments of \citet{matsuda2020_sup} and \citet{jananji2017_sup} to measure $\Kd$, anillin may be binding one or more actin filaments, and so it is not a true measure of $\Kd$, which is specifically for anillin binding in the overlap between two actin filaments.
The measured dissociation constant is likely an overestimate of the true $\Kd$ (i.e.~an underestimate of the binding affinity).
\citet{kucera2021_sup} indeed report a much smaller value of \SI{7.82}{\nano\molar} for $\Ks$, which was measured by imaging filaments and crosslinkers that were fluorescently labeled.
Because their method is more direct and specifically measures $\Ks$, we weight this value much more heavily than the other two reported values.
$\Kd$ will be smaller than $\Ks$, but without any data on how much more strongly the crosslinkers bind in an overlap region, we chose to use as a first estimate for $\Kd$ the same ratio of the dissociation constants of Ase1 between a single microtubule and in an overlap region, $\Kd = \nicefrac{\Ks}{5}$.

Unbinding rates are known to be affected by the application of force to one of the binding partners, thus in principle making the dissociation constant a function of the ring-constriction force~\cite{evans1997_sup}.
With an energy landscape reducible to a 1D order parameter with a single transition state, a very simple model of the force dependence of the unbinding rate assumes a constant force that tilts the energy landscape, but without changing the relative distance between the well of the bound state and the transition state~\cite{evans1997_sup,bell1978_sup}.
With this model, the force-dependent off-rate can be written as
\begin{equation}
    k_\mathrm{off}(F) = k_\mathrm{off}(0) \exp \mleft( \frac{F x_\beta}{\kb T} \mright),
    \label{eq:bell}
\end{equation}
where $F$ is the applied force, $k_\mathrm{off}(0)$ is the off-rate with no applied force, and $x_\beta$ is the distance from the well to the transition state.
To get a rough idea of the magnitude of this effect on the constriction force calculated with our model, we can use the force near the maximum ring radius with $\Nsca = 2$, $\Nf = 8$, and $\Lf = \SI{3.0}{\micro\meter}$, which is around \SI{3}{\pico\newton}.
A value for $k_\mathrm{off}(0)$ of anillin binding to a single actin filament is reported in Ref.~\cite{kucera2021_sup} (\SI{0.06}{\per\second}).
While $x_\beta$ is not known, it is likely on the order of a few angstroms~\cite{jiang2002_sup}.
If we take the worst-case scenario and set $x_\beta = \SI{1}{\nano\meter}$, and assume the crosslinkers experience the full ring tension that results from the constriction force of \SI{3}{\pico\newton}, \SI{0.5}{\pico\newton}, (rather than the tension divided by the number of crosslinkers), then the dissociation constant only marginally increases, from $\Ks = \SI{7.82}{\nano\molar}$ in the absence of any force to $\Ks(F) = \SI{8.77}{\nano\molar}$.
Plugging this back into \namecref{eq:bending-energy} \ref{eq:tracks-force} of the main text, the constriction force changes by only 2\%, even in this worst-case scenario.
Combined with the fact that a decrease in the constriction force can essentially be compensated for by shifting the crosslinker concentration, we chose not to include this effect in our model, which would have significantly increased its complexity, as it would have to be solved iteratively.


\appendix*

\section*{Appendix}

Starting from the same assumptions as the main text, we give an alternative derivation for the relationship between the friction coefficient for ring constriction $\zetaR$ and the friction coefficient of a single overlap $\zeta$.
To begin, consider the dynamics of the total overlap in the ring,
\begin{equation}
    \mathrm{d} L_\mathrm{tot} = \mleft[ \frac{F_\mathrm{tot}}{\zeta_\mathrm{tot}} + \deriv{}{\Ltot} \mleft( \frac{\kb T}{\zeta_\mathrm{tot}} \mright) \mright] \mathrm{d} t + \sqrt{ \frac{2 \kb T}{\zeta_\mathrm{tot}}} \mathrm{d} W, \label{eq:eqmLtot}
\end{equation}
where $F_\mathrm{tot}$ is the total linear force on the overlaps and $\zeta_\mathrm{tot}$ is the friction coefficient for the change in the total overlap.
If we approximate the dynamics of the total overlap as the sum of the dynamics of each individual overlap, then
\begin{subequations}
\begin{align}
    \deriv{\Ltot}{t} &= \sum_i^{\No} \deriv{L_i}{t}\\
                     &= \sum_i^{\No} \frac{F_i}{\zeta_i} \label{eq:independent-overlaps}\\
                     &= \frac{1}{\zeta} \sum_i^{\No} F_i\\
                     &= \frac{F_\mathrm{tot}}{\zeta},
\end{align}
\end{subequations}
where $\zeta_i$ is the friction coefficient of overlap $i$, but we assume $\zeta_i = \zeta$, i.e.~the friction coefficient is the same for all overlaps, which is reasonable if they have the same amount of overlap.
Having identified that $\zeta_\mathrm{tot} = \zeta$, we can than apply the chain rule and \namecref{eq:bending-energy} \ref{eq:connector} of the main text to write the radial ring-constriction dynamics in terms of $\zeta$,
\begin{subequations}
\begin{align}
    \deriv{R}{t} &= \deriv{R}{L} \deriv{L}{\Ltot} \deriv{\Ltot}{t}\\
                 &= -\frac{\Nsca}{2 \pi} \frac{1}{\No} \frac{F_\mathrm{tot}}{\zeta}\\
                 &= \frac{\Nsca}{2 \pi \No \zeta} \deriv{U}{R} \deriv{R}{L}\\
                 &= \frac{\Nsca^2}{4 \pi^2 \No} \frac{\ForceR}{\zeta},
\end{align}
\end{subequations}
which implies
\begin{equation}
    \zetaR = \frac{4 \pi^2 \No}{\Nsca^2} \zeta.
\end{equation}
We thus recover the same expression for $\zetaR$ as derived in \cref{eq:zetaR}.

Assuming a ring with perfectly circular geometry implies that a change in the radius leads to a simultaneous change in all the overlaps in the ring.
Such a mechanism would imply a single collective jump of all involved overlaps, which would require the crossing of a much higher free-energy barrier.
In reality, the ring is not perfectly circular, and the filaments will be locally deformed.
With local deformation, paths to filament sliding and ring constriction are possible in which only a single overlap changes at a time.
The approximation made in \cref{eq:independent-overlaps} is equivalent to assuming that the dynamics can be described with only these paths and that the local deformations are small enough to be ignored.

    \putbib[tex/supplemental,tex/main_prlNotes]
    \onecolumngrid
    \clearpage
    \pagebreak
    \section*{Supplemental figures}

\begin{figure}[!h]
    \includegraphics{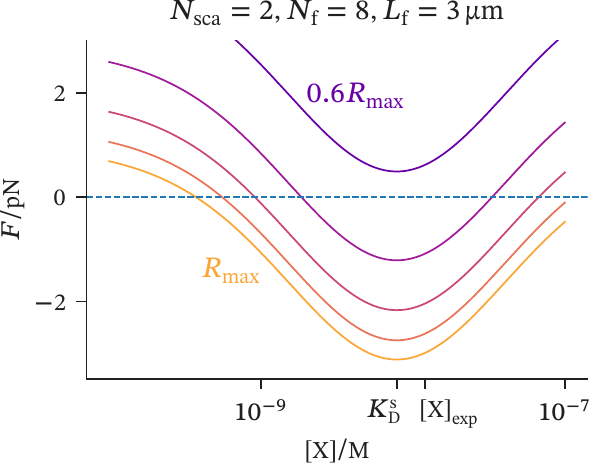}
    \caption{
        Semi-log plot of the radial constriction force as a function of the anillin concentration with $\Nf = 8$ and $\Nsca = 2$.
        The different coloured curves correspond to different values of $R$, expressed in the plot as a fraction of the maximal constriction, ranging from $0.6 R_\mathrm{max}$ to $R_\mathrm{max}$ in increments of $0.1 R_\mathrm{max}$.
        The constriction force is maximal at $\ce{[X]} = \Ks$, which is marked on the plot, along with the experimental crosslinker concentration $\ce{[X]}_\mathrm{exp}$~\cite{kucera2021_sup}.
        Data was produced with the parameters given in \cref{tab:params}.
        \label{fig:force-concentration}
     }
\end{figure}

\begin{figure}[!h]
    \centering
    \includegraphics{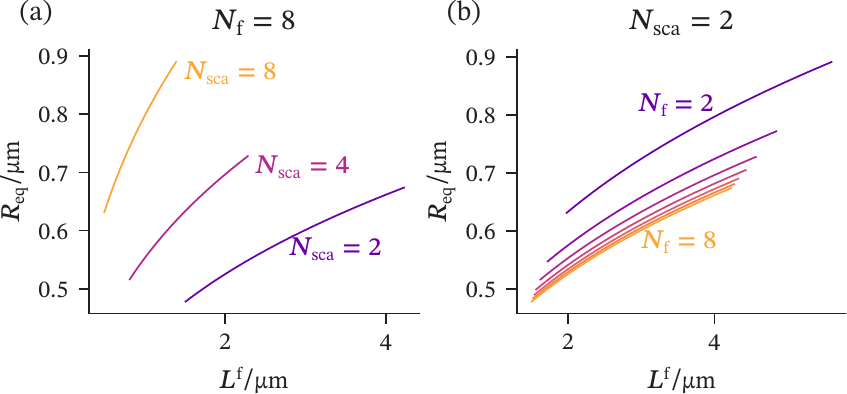}
    \caption{
        Equilibrium radius as a function of the filament length.
        (a) $\Req$ is plotted for three values of $\Nsca$, with $\Nf = 8$.
        (b) $\Req$ is plotted for a range of $\Nf$ values, with $\Nsca = 2$.
        To understand the end points of the curves, we must consider \namecref{fig:crosssection} \ref{fig:energy-force}(b)(v) of the main text: at smaller filament lengths, $R_\mathrm{eq}$ (the blue stars) approaches $R_\mathrm{max}$, while at larger filament lengths, $R_\mathrm{eq}$ approaches $R_\mathrm{min}$.
        Thus, the left endpoints, where $\Lf$ is small, occur when the bending force equals the sliding force at $R_\mathrm{max}$, while the right end points, where $\Lf$ is large, occur when the two forces are instead equal at $R_\mathrm{min}$.
        Data for all plots was produced with the parameters given in \cref{tab:params}.
        \label{fig:Req}
    }
\end{figure}

\begin{figure*}
    \centering
    \includegraphics{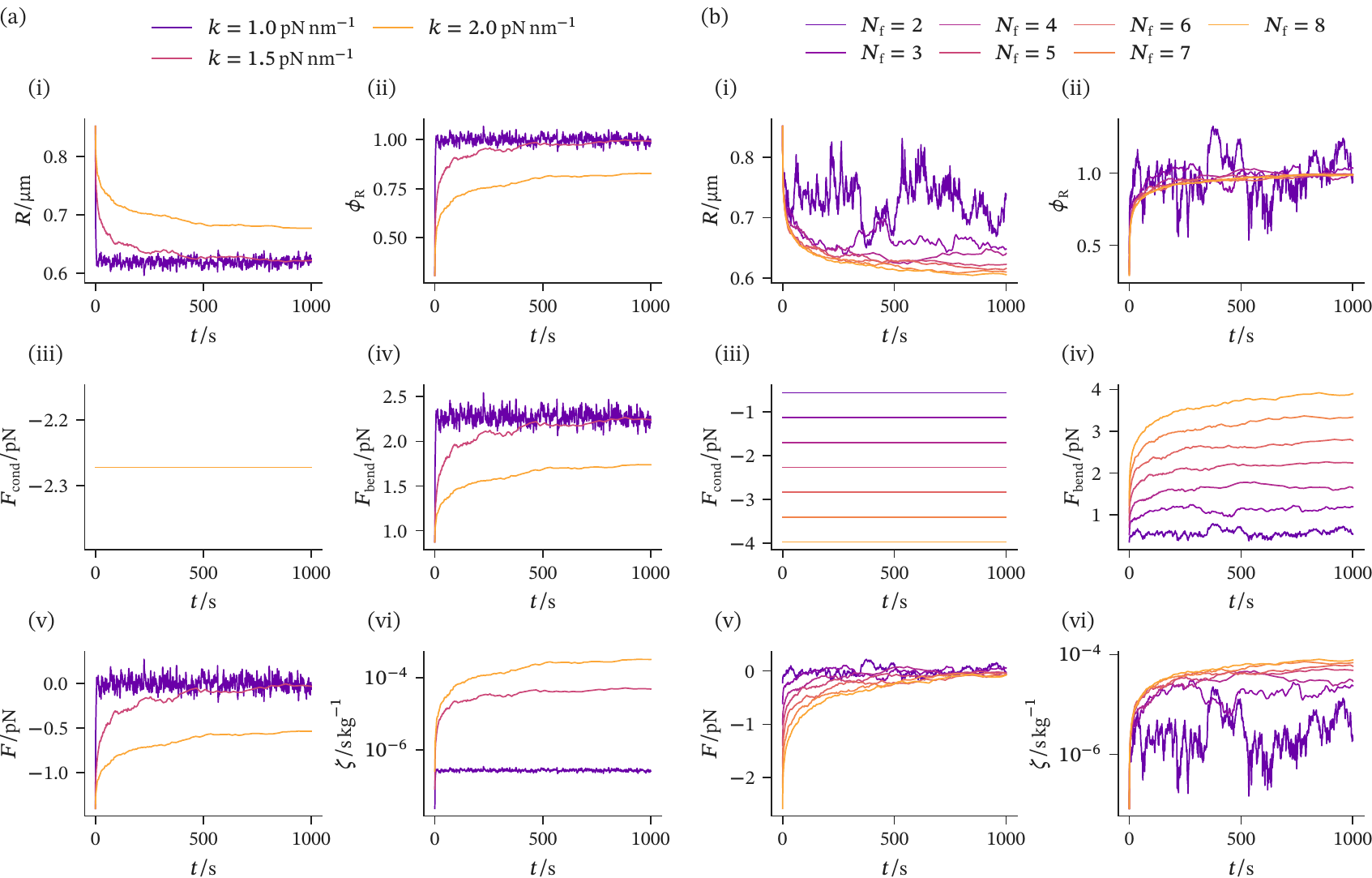}
    \caption{
        \label{fig:dynamics-sup}
        Ring constriction trajectories with $\Lf = \SI{3.0}{\micro\meter}$.
        In both (a) and (b), the plots are of (i) the ring radius $R$, (ii) the progress of radial constriction from its maximum ($\phi = 0$) to its equilibrium value ($\phi = 1$), where $\phi_\mathrm{R} = \nicefrac{\mleft( \Rmax - R \mright)}{\mleft( \Rmax - \Req \mright)}$; (iii) the condensation force $F_\mathrm{cond}$, (iv) the bending force $F_\mathrm{bend}$, (v) the total force $F$, and (vi) the friction coefficient of an overlap in the ring $\zeta$.
        (a) Ring constriction trajectories with $\Nsca = 2$, $\Nf = 5$, and three values of $k$.
        The ring is able to reach the equilibrium radius for a range of values under $k = \SI{2.0}{\pico\newton\per\nano\meter}$, although going as low as $k = \SI{1.0}{\pico\newton\per\nano\meter}$ produces dynamics faster than observed in the experiments.
        (b) Ring constriction trajectories with $\Nsca = 2$, $k = \SI{1.5}{\pico\newton\per\nano\meter}$ and $\Nf$ ranging from 2 to 8.
        With $\Nf \leq 3$, the noise is large, but otherwise, the noise is small, and the dynamics are largely similar for different values of $\Nf$.
        The dynamics were solved with an adaptive, stability-optimized stochastic Runge-Kutta method~\cite{rackauckas2017_sup,rackauckas2017b_sup,rackauckas2018_sup} using the parameters given in \cref{tab:params}.
        The initial state is $\ell = 10$, and samples are plotted at \SI{1}{\second} intervals.
    }
\end{figure*}

\begin{figure*}
    \centering
    \includegraphics{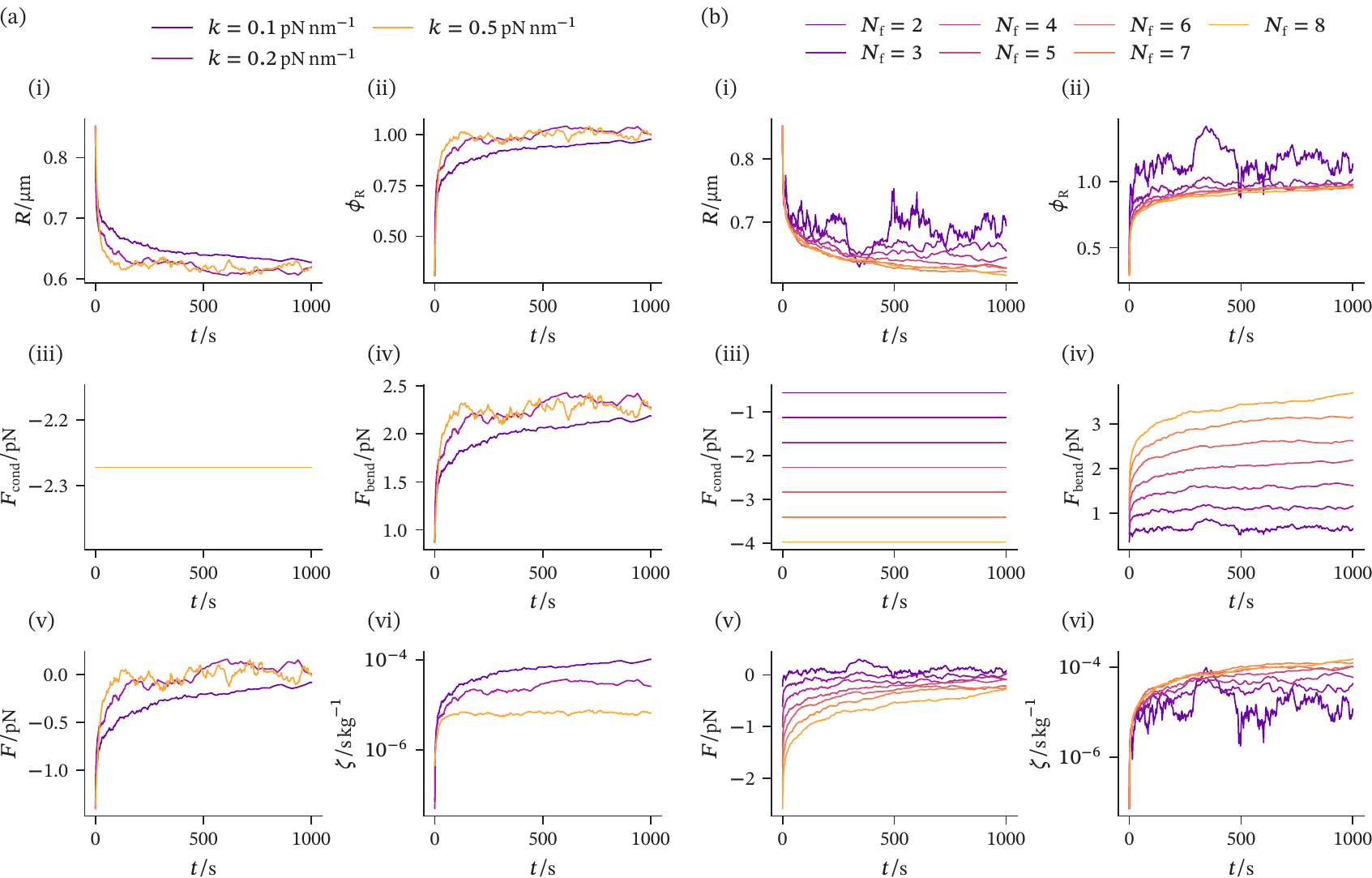}
    \caption{
        \label{fig:dynamics-small-k}
        Ring constriction trajectories with $\Lf = \SI{3.0}{\micro\meter}$ and values of $k$ that lead to $B < 0$ (\cref{eq:B}).
        For these values of $k$, the free-energy barrier is located at the in-register state that at higher $k$ is the free-energy trough.
        The quantities plotted in (i)--(vi) of both (a) and (b) are the same as in \cref{fig:dynamics-sup}.
        (a) Ring constriction trajectories with $\Nsca = 2$, $\Nf = 5$, and three values of $k$.
        The ring is able to reach the equilibrium radius in the experimental timescale~\cite{kucera2021_sup} for values of $k$ at least as small as $\SI{0.1}{\pico\newton\per\nano\meter}$.
        (b) Ring constriction trajectories with $\Nsca = 2$, $k = \SI{0.1}{\pico\newton\per\nano\meter}$ and $\Nf$ ranging from 2 to 8.
        As in \cref{fig:dynamics-sup}(b), with $\Nf \leq 3$, the noise is large, but otherwise, the noise is small, and the dynamics are largely similar for different values of $\Nf$.
        See \cref{fig:dynamics-sup} for numerical details.
    }
\end{figure*}

\begin{figure*}
    \centering
    \includegraphics{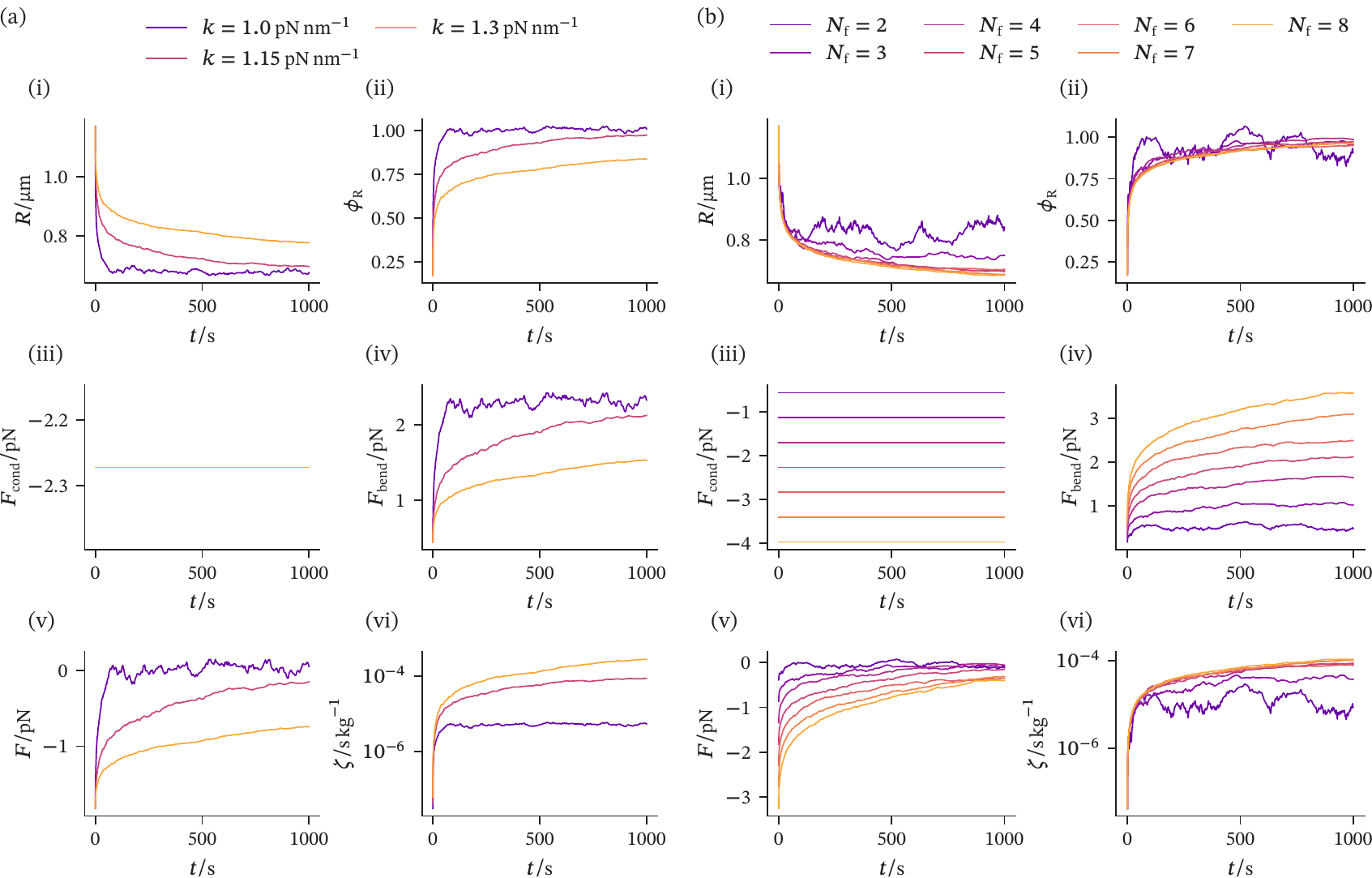}
    \caption{
        \label{fig:dynamics-Lf-4}
        Ring constriction trajectories with $\Lf = \SI{4.0}{\micro\meter}$.
        The quantities plotted in (i)--(vi) of both (a) and (b) are the same as in \cref{fig:dynamics-sup}.
        (a) Ring constriction trajectories with $\Nsca = 2$, $\Nf = 5$, and three values of $k$.
        In comparison to simulations with a smaller value of $\Lf$ (\cref{fig:dynamics-sup}), the ring constriction stalls at a lower value of $k$.
        (b) Ring constriction trajectories with $\Nsca = 2$, $k = \SI{1.15}{\pico\newton\per\nano\meter}$ and $\Nf$ ranging from 2 to 8.
        In comparison to simulations with a smaller value of $\Lf$ (\cref{fig:dynamics-sup}), the noise with $\Nf = 2$ is significantly lower.
        See \cref{fig:dynamics-sup} for numerical details.
    }
\end{figure*}

\begin{figure}
    \centering
    \includegraphics{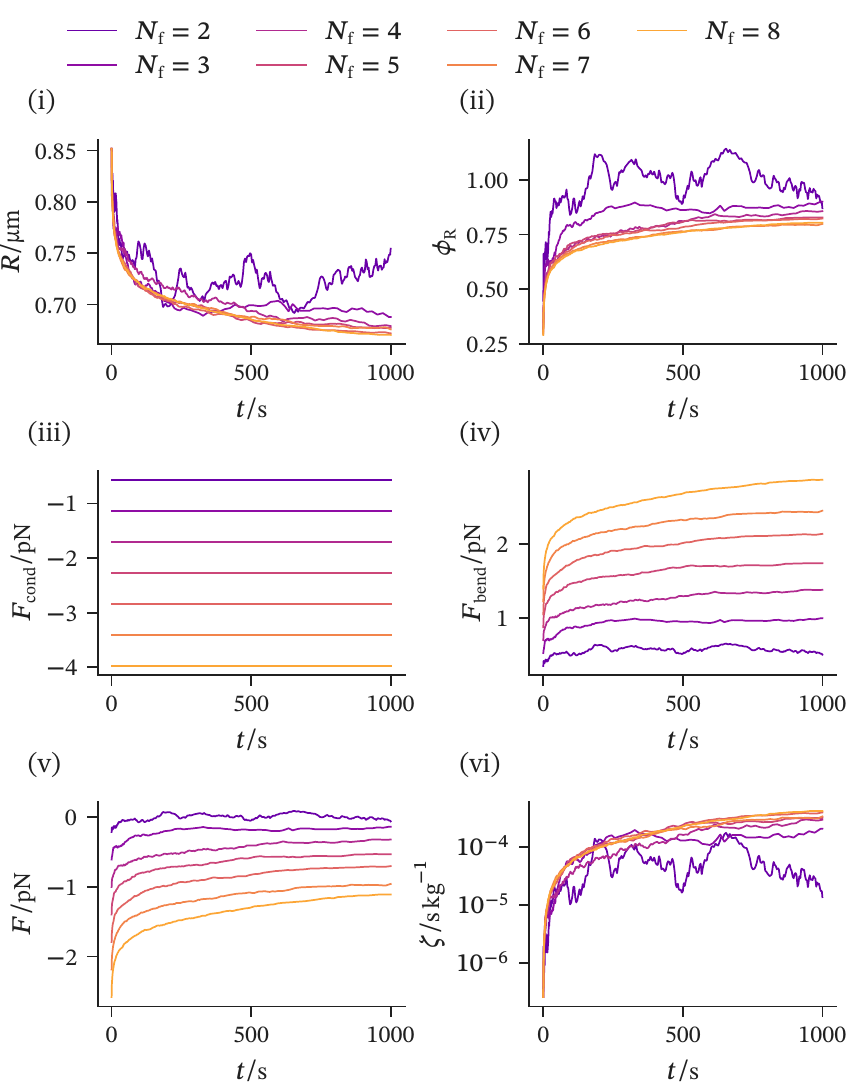}
    \caption{
        \label{fig:dynamics-k-2}
        Ring constriction trajectories with $\Lf = \SI{3.0}{\micro\meter}$, $\Nsca = 2$, $k = \SI{2.0}{\pico\newton\per\nano\meter}$, and $\Nf$ ranging from 2 to 8.
        The quantities plotted in (i)--(vi) are the same as in \cref{fig:dynamics-sup}.
        With the exception of when $\Nf = 2$, the constriction stalls at about the same radius for different values of $\Nf$.
        See \cref{fig:dynamics-sup} for numerical details.
    }
\end{figure}

    \end{bibunit}
\end{document}